\documentclass[aps, prd, twocolumn, nofootinbib,]{revtex4-2}

\usepackage{amsmath}
\usepackage{amsfonts}
\usepackage{wasysym}
\usepackage{graphicx}
\usepackage{comment}
\usepackage{tensor}
\usepackage{cancel}
\usepackage[svgnames]{xcolor}

\usepackage[colorlinks = true, linkcolor = DarkBlue, citecolor = DarkBlue, urlcolor = DarkBlue, bookmarks = false]{hyperref}

\def\filetype{pdf}
\def\path{}

\allowdisplaybreaks[1]

\begin{document}


\title{Einstein-Dirac-Maxwell wormholes in quantum field theory}
\author{Ben Kain}
\affiliation{Department of Physics, College of the Holy Cross, Worcester, Massachusetts 01610, USA}

\begin{abstract}
\noindent
Charged Dirac fields minimally coupled to gravity have spherically symmetric wormhole solutions known as Einstein-Dirac-Maxwell (EDM) wormholes.  EDM wormholes do not make use of exotic matter and exist in asymptotically flat general relativity.  We construct static spherically symmetric EDM wormhole configurations in quantum field theory using semiclassical approximations for gravity and the electromagnetic field.  Our framework is able to describe a broader class of EDM wormholes than previously considered and, being constructed in quantum field theory, puts EDM wormholes on firmer theoretical ground.
\end{abstract} 

\maketitle


\section{Introduction}

Einstein-Dirac-Maxwell (EDM) wormholes are formed from charged Dirac fields minimally coupled to gravity \cite{Blazquez-Salcedo:2020czn, Blazquez-Salcedo:2021udn, Konoplya:2021hsm}.  EDM wormholes are particularly noteworthy because they do not make use of exotic matter and exist in asymptotically flat general relativity.  The original static solutions found in \cite{Blazquez-Salcedo:2020czn} have some concerning properties \cite{Danielson:2021aor} which the asymmetric static solutions found in \cite{Konoplya:2021hsm} avoid.  The asymmetric static solutions were numerically evolved forward in time using a time dependent EDM wormhole model in \cite{Kain:2023ann}.  In \cite{kain_EREPR}, the time evolution of EDM wormholes was used as a concrete model for the $\text{ER} = \text{EPR}$ conjecture \cite{Maldacena:2013xja}.  Additional works on EDM wormholes include \cite{Bolokhov:2021fil, Stuchlik:2021guq, Churilova:2021tgn, Wang:2022aze}.

These studies of EDM wormholes make use of two independent Dirac fields and fix the total angular momentum to $j = 1/2$. If a one-particle restriction is implemented for each Dirac field, then the fields can be interpreted as first quantized wave functions, while gravity and the electromagnetic field are treated classically.  This setup was first used for non-wormhole star like systems (see, for example, \cite{Finster:1998ws, Finster:1998ux, Herdeiro:2017fhv, Dzhunushaliev:2018jhj, Daka:2019iix}), before being applied to a wormhole geometry.

In our recent study of the non-wormhole Einstein-Dirac system \cite{Kain:2023jgu}, we used a single Dirac field minimally coupled to gravity and constructed star like configurations in quantum field theory using the semiclassical gravity approximation.  In place of multiple independent Dirac fields, we allowed for multiple excitations of a single quantized field.  The star like configurations were therefore populated with identical quantum spin-1/2 fermions.

The goal of this paper is to make a similar study of the EDM wormhole system.  With respect to \cite{Kain:2023jgu}, this requires generalizing the spacetime metric so that it can accommodate a wormhole, introducing the electromagnetic field, and charging the Dirac field.  We choose to quantize the Dirac field in the background of curved spacetime and the electromagnetic field, which means that we do not quantize the electromagnetic field.  With this simplification, we do not have to treat the electromagnetic interaction perturbatively.  We treat gravity and the electromagnetic field semiclassically by sourcing them with expectation values of the stress-energy-momentum tensor and the electromagnetic current.

Our framework improves the theoretical standing of EDM wormholes.  Most wormhole solutions constructed in the literature depend on exotic matter, make use of modifications to general relativity, or exist in anti--de Sitter space.  As mentioned above, EDM wormholes are noteworthy in that they are constructed using standard matter in asymptotically flat general relativity.  That they can now be described within quantum field theory and with semiclassical approximations puts EDM wormholes on firmer theoretical ground, which is advantageous given the hypothetical nature of wormholes.

Solutions to the system of equations we derive describe spherically symmetric EDM wormholes. These wormholes can have multiple values of the quantum numbers $n$ and $j$ excited, where $n$ counts the number of radial nodes and $j$ labels the total angular momentum, and can be populated by particles or antiparticles.  Our framework therefore describes a broader range of EDM wormholes than has previously been considered.  

In Sec.~\ref{sec:EDM}, after describing the metric and Lagrangian for our model, we quantize the Dirac field and present the semiclassical Einstein and electromagnetic field equations.  In Sec.~\ref{sec:states}, we construct spherically symmetric excitations of the quantized Dirac field, present the stress-energy-momentum tensor, and evaluate expectation values.  In Sec.~\ref{sec:EDM wormholes}, we discuss our numerical methods for solving the system of equations and present example EDM wormhole configurations.   We conclude in Sec.~\ref{sec:conclusion}.  Throughout we use units such that $c = \hbar = 1$.  


\section{Einstein-Dirac-Maxwell system}
\label{sec:EDM}

We study static spherically symmetric EDM wormholes.  We use the same form for the static spherically symmetric metric that is used in  \cite{Calhoun:2022xrw, Kain:2023ann},
\begin{equation} \label{metric}
ds^2 = -\alpha^2(r) dt^2 + A(r)dr^2 + C(r) (d\theta^2 + \sin^2\theta d\phi^2),
\end{equation}
where $-\infty < r < \infty$.  Following \cite{Calhoun:2022xrw, Kain:2023ann}, we define 
\begin{equation}
R(r) \equiv \sqrt{C(r)}
\end{equation}
as the areal radius.  Since the minimum of the areal radius occurs at $r = 0$ for EDM wormholes, we define
\begin{equation} \label{R0 def}
R_0 \equiv R(0)
\end{equation}
as the radius of the wormhole throat and interpret positive and negative $r$ as the spatial regions for the two sides of the wormhole.  When studying wormholes, it is necessary to parametrize the metric such that it can accommodate a nonzero wormhole throat radius.  It is for this reason that we include the metric function $C(r)$ in Eq.~(\ref{metric}).  Including $C(r)$ and allowing $r$ to take positive and negative values are the principle differences between the metric we use here and the metric we used in our recent study of the Einstein-Dirac system in \cite{Kain:2023jgu}, where we set $C = r^2$ and required $r\geq 0$.
 
For the EDM system, we include a single charged Dirac spinor field, $\psi$, and the electromagnetic field, $\mathcal{A}_\mu$, both minimally coupled to gravity.  The Lagrangian is
\begin{equation} \label{Lagrangian}
\mathcal{L} = \sqrt{-\det(g_{\mu\nu})} 
\left(\frac{\mathcal{R}}{16\pi G} + \mathcal{L}_\psi + \mathcal{L}_\mathcal{A}\right),
\end{equation}
where $\mathcal{R}$ is the Ricci scalar, $G$ is the gravitational constant, and
\begin{equation}
\begin{split}
\mathcal{L}_\psi &= \bar{\psi} \gamma^\mu D_\mu \psi - m_\psi \bar{\psi} \psi
\\
\mathcal{L}_\mathcal{A} &= -\frac{1}{4} F_{\mu\nu} F^{\mu\nu}
\\
D_\mu \psi &= \nabla_\mu \psi - ie \mathcal{A}_\mu \psi 
\\
F_{\mu\nu} &= \partial_\mu \mathcal{A}_\nu - \partial_\nu \mathcal{A}_\mu ,
\end{split}
\end{equation}
where $m_\psi$ is the Dirac mass parameter, $e$ is the electric charge, and $F_{\mu\nu}$ is the field strength.  Conventions and details for the spinor field, such as the definitions of the covariant derivative and the adjoint spinor, are the same as those in \cite{Kain:2023jgu}, to which we refer the reader.  We use the following vierbein to couple the spinor to spacetime,
\begin{equation} \label{vierbein}
\gamma^t = \frac{\tilde{\gamma}^t}{\alpha},
\quad
\gamma^r = \frac{\tilde{\gamma}^r}{\sqrt{A}},
\quad
\gamma^\theta = \frac{\tilde{\gamma}^\theta}{\sqrt{C}}, 
\quad
\gamma^\phi = \frac{\tilde{\gamma}^\phi}{\sqrt{C} \sin\theta},
\end{equation}
where the $\gamma^\mu$ are curved space $\gamma$-matrices and the $\tilde{\gamma}^\mu$ are flat space $\gamma$-matrices defined by
\begin{equation} \label{vierbein gamma}
\begin{split}
\tilde{\gamma}^t &= \tilde{\gamma}^0
\\
\tilde{\gamma}^r &= \tilde{\gamma}^1 \sin\theta \cos\phi + \tilde{\gamma}^2 \sin\theta \sin\phi + \tilde{\gamma}^3 \cos\theta
\\
\tilde{\gamma}^\theta &= \tilde{\gamma}^1 \cos\theta \cos\phi + \tilde{\gamma}^2 \cos\theta \sin\phi - \tilde{\gamma}^3 \sin\theta
\\
\tilde{\gamma}^\phi &= -\tilde{\gamma}^1\sin\phi + \tilde{\gamma}^2\cos\phi.
\end{split}
\end{equation}
We use the Dirac representation for the flat space $\gamma$-matrices
\begin{equation} \label{Dirac rep}
\tilde{\gamma}^0 = i
\begin{pmatrix}
1 & 0 \\
0 & -1
\end{pmatrix},
\qquad
\tilde{\gamma}^j = i
\begin{pmatrix}
0 & \sigma^j \\
-\sigma^j & 0
\end{pmatrix},
\end{equation}
where $j=1,2,3$ and where
\begin{equation}
\sigma^1 = 
\begin{pmatrix}
0 & 1 \\ 1 & 0
\end{pmatrix},
\qquad
\sigma^2 = 
\begin{pmatrix}
0 & -i \\ i & 0
\end{pmatrix},
\qquad
\sigma^3 = 
\begin{pmatrix}
1 & 0 \\ 0 & -1
\end{pmatrix}
\end{equation}
are the Pauli matrices.  We note that the vierbein in Eq.~(\ref{vierbein}) differs from the vierbein used in \cite{Kain:2023jgu} only because the metrics differ, while the flat space $\tilde{\gamma}^\mu$ in Eq.~(\ref{vierbein gamma}) and the flat space $\tilde{\gamma}^a$ in Eq.~(\ref{Dirac rep}) are the same.  The flat and curved space $\gamma$-matrices obey
\begin{equation}
\{\gamma^\mu, \gamma^\nu\} = 2g^{\mu\nu},
\quad
\{\tilde{\gamma}^\mu, \tilde{\gamma}^\nu\} = 2\eta^{\mu\nu},
\quad
\{\tilde{\gamma}^a, \tilde{\gamma}^b\} = 2\eta^{ab},
\end{equation}
where $\eta^{\mu\nu} = \eta^{ab} = \text{diag}(-1,1,1,1)$ is the flat space metric.

In a spherically symmetric spacetime, the angular components of a spacetime vector must vanish, so that $\mathcal{A}_\mu = (\mathcal{A}_t, \mathcal{A}_r, 0, 0)$.  For a static spherically symmetric spacetime, $\psi = \psi(t,r,\theta,\phi)$ and $\mathcal{A}_\mu = \mathcal{A}_\mu(r)$.  We fix the $U(1)$ gauge such that
\begin{equation} \label{Ar = 0}
\mathcal{A}_r = 0.
\end{equation}
There is still some residual gauge freedom, which we use to set
\begin{equation} \label{At(0) = 0}
\mathcal{A}_t(0) = 0.
\end{equation}
With the gauge choice in (\ref{Ar = 0}), the only nonvanishing components of the field strength are
\begin{equation} \label{Ftr}
F_{tr} = -F_{rt} 
= -\partial_r \mathcal{A}_t.
\end{equation}

In the following subsections, we canonically quantize the charged Dirac field and present the semiclassical equations of motion for the gauge field and the semiclassical Einstein field equations.


\subsection{Dirac}
\label{sec:Dirac}

From the Lagrangian in (\ref{Lagrangian}), the classical equation of motion for the Dirac field is the charged Dirac equation,
\begin{equation} \label{spinor eom}
\gamma^\mu (\nabla_\mu - ie \mathcal{A}_\mu)f_I - m_\psi f_I
= 0,
\end{equation}
where we have labeled the solution as $f_I$ instead of as $\psi$.  The $f_I$, where $I$ represents some set of quantum numbers, are classical solutions to the equations of motion.  We will use the $f_I$ as mode functions in the expansion of the Dirac field operator when we quantize the Dirac field below.

It is straightforward to show that if the mode function $f_I$ satisfies the equations of motion, then the charge conjugated mode function $f_I^c \equiv \tilde{\gamma}^2 f_I^*$ satisfies the equations of motion with the opposite sign charge,
\begin{equation} 
\gamma^\mu \left[\nabla_\mu - i(-e) \mathcal{A}_\mu\right]f_I^c - m_\psi f_I^c
= 0.
\end{equation}
In flat space, we can interpret this to mean that if $f_I$ describes a particle, $f_I^c$ describes the antiparticle.  Since our metric is static, this interpretation continues to hold in curved space.  We can therefore write the equations of motion as
\begin{equation} \label{classical eom}
\gamma^\mu (\nabla_\mu \mp ie \mathcal{A}_\mu)f_I^\pm - m_\psi f_I^\pm
= 0.
\end{equation}
In this form, if $f_I^+$ describes a particle, then $f_I^-$ describes the antiparticle.

The equations of motion in (\ref{classical eom}) may be further rewritten into the form \cite{Kain:2023jgu}
\begin{equation} \label{H eom}
\widehat{H}_\pm f_I^\pm = i \partial_t f_I^\pm,
\end{equation}
where
\begin{equation} \label{Hhat def}
\begin{split}
\widehat{H}_\pm
&\equiv 
\frac{i\alpha}{\sqrt{A}} \tilde{\gamma}^0 \tilde{\gamma}^r\left(\partial_r  + \frac{\partial_r \alpha}{2\alpha} + \frac{\partial_r C}{2C} \right) 
- \frac{\alpha}{\sqrt{C}} \tilde{\gamma}^r
\widehat{K} 
\\
&\qquad
- i\alpha m_\psi \tilde{\gamma}^0 
\mp e  \mathcal{A}_t.
\end{split}
\end{equation}
The operator $\widehat{K}$ is defined by $\widehat{K} = -i\tilde{\gamma}^0 (\widehat{J}^{\,2} - \widehat{L}^2 + 1/4)$, where $\widehat{J}^{\,2}$ and $\widehat{L}^2$ are the standard operators for total and orbital angular momentum.  In Sec.~IV of \cite{Kain:2023jgu}, we gave a detailed description of solving the uncharged Dirac equation for the corresponding mode functions.  Generalizing this procedure for the metric in (\ref{metric}) and for the inclusion of the gauge field, the classical solutions to the charged Dirac equation are
\begin{equation} \label{mode functions}
\!
f_{njm_j\pm}(t,r,\theta,\phi)
= 
\frac{e^{\mp i\omega_{nj} t}}{\sqrt{\alpha(r) C(r)}}
\begin{pmatrix}
\hphantom{i} P_{nj\pm}(r) \mathcal{Y}_{j\mp1/2}^{m_j} (\theta,\phi)
\\[4pt]
i P_{nj\mp}^*(r) \mathcal{Y}_{j\pm1/2}^{m_j} (\theta,\phi)
\end{pmatrix}.
\end{equation}
These mode functions depend on the radial quantum number $n=0,1,2,\ldots$, the total angular momentum quantum number $j=1/2, 3/2, 5/2, \ldots$, the three-component of total angular momentum quantum number $m_j = -j, -j+1, \ldots, j-1,j$, and the quantum number $\pm$ which indicates whether the solution describes a particle or an antiparticle.  The $\omega_{nj}$ are real constants and the $\mathcal{Y}_{j\pm1/2}^{m_j}$ are two-component spin angle functions given by \cite{Sakurai, Bransden}
\begin{equation} \label{spin angle}
\begin{split}
\mathcal{Y}_{j-1/2}^{m_j} &= 
\sqrt{\frac{j+m_j}{2j}} Y_{j-1/2}^{m_j-1/2}
\begin{pmatrix}
1 \\ 0
\end{pmatrix}
\\
&\qquad
+
\sqrt{\frac{j-m_j}{2j}} Y_{j-1/2}^{m_j+1/2}
\begin{pmatrix}
0 \\ 1
\end{pmatrix}
\\
\mathcal{Y}_{j+1/2}^{m_j} &= 
- \sqrt{\frac{j+1-m_j}{2j+2}} Y_{j+1/2}^{m_j-1/2}
\begin{pmatrix}
1 \\ 0
\end{pmatrix}
\\
&\qquad
+
\sqrt{\frac{j+1+m_j}{2j+2}} Y_{j+1/2}^{m_j+1/2}
\begin{pmatrix}
0 \\ 1
\end{pmatrix},
\end{split}
\end{equation}
where the $Y_\ell^{m_\ell}$ are spherical harmonics.  The $P_{nj\pm}$ are one-component radial functions.  They obey the radial equations of motion, 
\begin{equation} \label{radial eom}
\begin{split}
\partial_r P_{nj\pm}
&=
\mp \frac{\sqrt{A}}{\alpha}
\biggl[ \left( e\mathcal{A}_t + \omega_{nj} \pm  \alpha m_\psi \right)
P_{nj\mp}^* 
\\
&\qquad
- \frac{\alpha}{\sqrt{C}} 
\left(j+\frac{1}{2}\right)
P_{nj\pm}
\biggr],
\end{split}
\end{equation}
which follow from the equations of motion in (\ref{H eom}).  The radial equations of motion do not contain any factors of $i$.  This indicates that we can take the $P_{nj\pm}$ to be purely real, which we shall do from this point forward.  The radial equations of motion also do not depend on $m_j$ and, for this reason, we take the $P_{nj\pm}$ and the $\omega_{nj}$ to be independent of $m_j$.  Finally, the charge conjugated mode functions are given by  $f^c_{njm_j+} = i (-1)^{m_j+1/2} f_{n,j,-m_j,-}$, which differ by an irrelevant global phase from $f_{n,j,-m_j,-}$.

The frequencies $\omega_{nj}$ depend on the choice of $U(1)$ gauge.  However, if mode function $f_{njm_j+}$ has frequency $\omega_{nj}$, then, independent of the gauge choice, mode function $f_{njm_j-}$ has frequency $-\omega_{nj}$.  Classifying the mode functions as satisfying
\begin{equation} \label{pos sol}
\xi^\mu \partial_\mu f_{njm_j+} = \partial_t f_{njm_j+} =  -i\omega_{nj} f_{njm_j+},
\end{equation}  
or
\begin{equation} \label{neg sol}
\xi^\mu \partial_\mu f_{njm_j-} = \partial_t f_{njm_j-} =  +i\omega_{nj} f_{njm_j-},
\end{equation}
where $\xi^\mu = (1,0,0,0)$ is the hyperspace-orthogonal timelike Killing vector for the metric in (\ref{metric}) \cite{Birrell:1982ix, Carroll:2004st}, is then gauge invariant.  In the uncharged case, the classifications are labeled ``positive frequency" and ``negative frequency."  We avoid these labels here since, in the charged case, the sign of $\omega_{nj}$ can be changed with a gauge transformation.  Instead, the quantum number $\pm$ specifies the classification.  This classification allows us to identify a preferred vacuum state and to have a natural definition for particle and antiparticle.

Having found the mode functions, we now quantize the Dirac field in a background of curved spacetime and an electromagnetic field \cite{Birrell:1982ix, Wald:1995yp, Parker:2009uva}.  Neither spacetime nor the electromagnetic field will be quantized, but they will both be treated semiclassically, as we explain in the next two subsections.   We introduce an inner product that has the same form as used in \cite{Kain:2023jgu}, but now the inner product is defined on the space of solutions to the charged Dirac equation,
\begin{equation} \label{inner product}
(f_I, f_J) = \int_\Sigma d^3x \sqrt{\det(\gamma_{ij})}\, f_I^\dag f_J,
\end{equation}
where $\gamma_{ij}$ is the induced spatial metric on the spatial hypersurface $\Sigma$, $\det(\gamma_{ij}) = A C^2 \sin^2\theta$, and $d^3x = dr d\theta d\phi$.  Using the mode functions in (\ref{mode functions}) and the orthonormality of the spin angle functions,
\begin{equation} \label{orthonormal spin angle}
\int d\theta d\phi \sin\theta
\left(\mathcal{Y}_{j\pm1/2}^{m_j}\right)^\dag
\mathcal{Y}_{j'\pm'1/2}^{m_j'}
= \delta_{j, j'} \delta_{m_j, m_j'} \delta_{\pm, \pm'},
\end{equation}
we have
\begin{align} \label{(fI,fj)}
\begin{split}
&(f_{njm_j\pm}, f_{n'j'm_j'\pm'})
\\
&\qquad
= \delta_{j,j'} \delta_{m_j,m_j'} \delta_{\pm, \pm'}
e^{\pm i(\omega_{nj} - \omega_{n'j})t}
\\
&\qquad\qquad
\times\int_{-\infty}^\infty dr \frac{\sqrt{A}}{\alpha}
\left( 
P_{nj+} 
P_{n'j+}
+
P_{nj-} 
P_{n'j-}
\right).
\end{split}
\end{align}
When the mode functions have the same sign for their electric charge, the inner product can be written in terms of a current, $if_I^\pm \gamma^\mu f_J^\pm$, which is divergence-free, $\nabla_\mu(if_I^\pm \gamma^\mu f_J^\pm) = 0$.  Although the current is no longer divergence-free when the mode functions have opposite signs for their electric charge, the inner product in this case always vanishes, $(f_I^\pm, f_J^\mp) = 0$.  Assuming the mode functions decay sufficiently fast at spatial infinity, the inner product is then time independent, as required.  The equations of motion in (\ref{H eom}) become the eigenvalue equations $\widehat{H}_\pm f_I^\pm = \pm\omega_I f_I^\pm$ and $\widehat{H}_\pm$ can be shown to be Hermitian with respect to the inner product.  The inner product is then orthogonal with respect to the quantum number $n$ as long as the eigenvalues are distinct for different values of $n$.

From (\ref{(fI,fj)}), the norm of a mode function is given by
\begin{equation} \label{(fI,fI)}
\mathcal{N}_{nj} \equiv (f_I, f_I)
=
\int_{-\infty}^\infty dr \frac{\sqrt{A}}{\alpha}
\left( 
P_{nj+}^2
+
P_{nj-}^2
\right).
\end{equation}
Since the integrand is positive definite, $f_I$ can be scaled such that the norm is equal to $1$.  We explain in Sec.~\ref{sec:EDM wormholes} how we implement this scaling and we assume our mode functions are normalized,
\begin{equation} \label{norm}
\mathcal{N}_{nj} = 1.
\end{equation}
We have now established that the mode functions satisfy the orthonormality conditions
\begin{equation}
(f_I, f_J) = \delta_{IJ}.
\end{equation}

Having constructed an orthonormal set of mode functions, the Dirac field can be expanded in terms of them.  We then promote the field to an operator,
\begin{equation} \label{field decomposition}
\hat{\psi}(t,\vec{x}) = \sum_I \left[ \hat{b}_I f_I^+(t, \vec{x}) + \hat{d}^\dag_I f_I^-(t,\vec{x}) \right].
\end{equation}
The momentum conjugate to $\psi$ is given by
\begin{equation} \label{pi def}
\pi = \frac{\partial \mathcal{L}}{\partial(\partial_t \psi)} 
= \sqrt{\det(\gamma_{ij})} \, i \psi^\dag,
\end{equation}
which we also promote to an operator, $\hat{\pi}$.  We impose equal time anticommutation relations between the field operators,
\begin{equation} \label{canonical anticommuation}
\begin{split}
\{ \hat{\psi}_a(t,\vec{x}), \hat{\pi}_b(t,\vec{y}) \} &= i\delta_{ab} \delta^3(\vec{x} - \vec{y}),
\\
\{ \hat{\psi}_a(t,\vec{x}), \hat{\psi}_b(t,\vec{y}) \} &= 0
\\
\{ \hat{\pi}_a(t,\vec{x}), \hat{\pi}_b(t,\vec{y}) \} &= 0,
\end{split}
\end{equation}
where $a$ and $b$ label the Dirac spinor components.  Plugging the field decompositions for $\hat{\psi}$ and $\hat{\pi}$ into the anticommutation relations and using the orthonormality of the mode functions, we find the standard anticommutation relations for creation and annihilation operators,
\begin{equation} \label{b d anticommutation}
\{ \hat{b}_I, \hat{b}^\dag_J \} = \delta_{IJ},
\qquad
\{ \hat{d}_I, \hat{d}_J^\dag \} = \delta_{IJ},
\end{equation}
with all other anticommutators vanishing.


\subsection{Maxwell}
\label{sec:Maxwell}

From the Lagrangian in (\ref{Lagrangian}), the classical equations of motion for the gauge field are Maxwell's equations,
\begin{equation} \label{classical gauge eom}
\nabla_\mu F^{\mu\nu} = j^\nu, \qquad
j^\nu = ie \bar{\psi} \gamma^\nu \psi,
\end{equation}
where $j^\nu$ is the classical electromagnetic current.  To incorporate the quantized Dirac field, and in analogy to the semiclassical gravity approximation we use in the next subsection \cite{Birrell:1982ix, Wald:1995yp, Parker:2009uva, Hu:2020luk}, we use a semiclassical approximation for the gauge field equations of motion by sourcing them with the expectation value of the electromagnetic current,
\begin{equation} \label{gauge eom}
\nabla_\mu F^{\mu\nu} = \langle \hat{j}^\nu \rangle, \qquad
\hat{j}^\nu = ie  \hat{\bar{\psi}} \gamma^\nu \hat{\psi} .
\end{equation}
The electromagnetic field is described classically by the left-hand side of the semiclassical equations of motion, while a quantum description of the Dirac field is used on the right-hand side.

Inserting the field decomposition for the Dirac field operator in (\ref{field decomposition}) into the current operator in (\ref{gauge eom}), we find
\begin{equation} \label{jhat}
\begin{split}
\hat{j}^\nu &= \sum_{I,J} 
\biggl[
j^\nu(f_I^+, f_J^+)
\hat{b}^\dag_I \hat{b}_J 
+
j^\nu(f_I^+, f_J^-)
\hat{b}^\dag_I  \hat{d}^\dag_J 
\\
&\qquad
+
j^\nu(f_I^-, f_J^+)
\hat{d}_I  \hat{b}_J 
+ 
j^\nu(f_I^-, f_J^-)
\hat{d}_I\hat{d}^\dag_J 
\biggr],
\end{split}
\end{equation}
where
\begin{equation} \label{J(f,f)}
j^\nu (f_I, f_J) \equiv i e \bar{f}_I \gamma^\nu f_J.
\end{equation}
The expectation value of the electromagnetic current operator is divergent since the current operator contains products of the field operator.  This divergence must be regulated and renormalized.  Following \cite{Alcubierre:2022rgp, Kain:2023jgu}, we normal order the current operator,
\begin{equation} \label{gauge eom 2}
\nabla_\mu F^{\mu\nu} = \langle {:\mathrel{\hat{j}^\nu}:} \rangle, 
\end{equation}
where 
\begin{equation} \label{jhat 2}
\begin{split}
{:\mathrel{\hat{j}^\nu}:} &= \sum_{I,J} 
\biggl[
j^\nu(f_I^+, f_J^+)
\hat{b}^\dag_I \hat{b}_J 
+
j^\nu(f_I^+, f_J^-)
\hat{b}^\dag_I  \hat{d}^\dag_J 
\\
&\qquad
+
j^\nu(f_I^-, f_J^+)
\hat{d}_I  \hat{b}_J 
- 
j^\nu(f_I^-, f_J^-)
\hat{d}^\dag_J \hat{d}_I
\biggr],
\end{split}
\end{equation}
which leads to finite results.  A more sophisticated renormalization scheme would be interesting \cite{Birrell:1982ix, Wald:1995yp, Parker:2009uva}, but is beyond the scope of this work.

With the $U(1)$ gauge condition in (\ref{Ar = 0}), the only nonvanishing gauge field component is $\mathcal{A}_t$.  From (\ref{gauge eom 2}), the equations of motion for $\mathcal{A}_t$ work out to
\begin{equation} \label{gauge field eom}
\partial_r^2 \mathcal{A}_t
= 
\left( \frac{\partial_r \alpha}{\alpha} +  \frac{\partial_r A}{2A} - \frac{\partial_r C}{C} \right) \partial_r \mathcal{A}_t
- 
\alpha^2 A
\langle {:\mathrel{\hat{j}^t}:} \rangle.
\end{equation}


\subsection{Einstein}
\label{sec:Einstein}

From the Lagrangian in (\ref{Lagrangian}), the classical equations of motion for gravity are the Einstein field equations,
\begin{equation}
G_{\mu\nu} = 8\pi G T_{\mu\nu},
\end{equation}
where $G_{\mu\nu}$ is the Einstein tensor and $T_{\mu\nu}$ is the classical stress-energy-momentum tensor,
\begin{equation}
T_{\mu\nu} = T^\psi_{\mu\nu}  + T^\mathcal{A}_{\mu\nu},
\end{equation}
where
\begin{equation} \label{SEMT}
\begin{split}
T^\psi_{\mu\nu} 
&=
-\frac{1}{4} \biggl[
\bar{\psi} \gamma_\mu D_\nu \psi
+ \bar{\psi} \gamma_\nu D_\mu \psi
\\
&\qquad 
- (D_\mu \bar{\psi}) \gamma_\nu \psi
- (D_\nu \bar{\psi}) \gamma_\mu \psi
\biggr]
\\
T^\mathcal{A}_{\mu\nu}
&=
g^{\alpha\beta}F_{\mu\alpha} F_{\nu\beta}
- \frac{1}{4} g_{\mu\nu} F_{\alpha\beta} F^{\alpha\beta}.
\end{split}
\end{equation}
To incorporate the quantized Dirac field, we use the semiclassical gravity approximation \cite{Birrell:1982ix, Wald:1995yp, Parker:2009uva, Hu:2020luk} and source the Einstein field equations with the expectation value of the Dirac field operator's contribution to the stress-energy-momentum tensor while keeping the gauge field's contribution classical,
\begin{equation} \label{EFE}
G_{\mu\nu} = 8\pi G 
\left( \langle \widehat{T}^\psi_{\mu\nu} \rangle + T^\mathcal{A}_{\mu\nu} \right),
\end{equation}
where
\begin{equation} \label{<Tpsi>}
\begin{split}
\widehat{T}^\psi_{\mu\nu} 
&= -\frac{1}{4} \biggl[
\hat{\bar{\psi}} \gamma_\mu D_\nu \hat{\psi}
+ \hat{\bar{\psi}} \gamma_\nu D_\mu \hat{\psi}
\\
&\qquad
- (D_\mu \hat{\bar{\psi}}) \gamma_\nu \hat{\psi}
- (D_\nu \hat{\bar{\psi}}) \gamma_\mu \hat{\psi}
\biggr].
\end{split}
\end{equation}
Gravity is described classically by the left-hand side of the semiclassical Einstein field equations, while a quantum description of the Dirac field is used on the right-hand side.

Inserting the field decomposition for the Dirac field operator in (\ref{field decomposition}) into the  stress-energy-momentum tensor operator in (\ref{<Tpsi>}), we find
\begin{equation} \label{hat T}
\begin{split}
\widehat{T}^\psi_{\mu\nu}
&= \sum_{I,J} \biggl[
T^\psi_{\mu\nu}(f_I^+, f_J^+)
\hat{b}^\dag_I \hat{b}_J 
+ 
T^\psi_{\mu\nu}(f_I^+, f^-_J)
\hat{b}^\dag_I \hat{d}^\dag_J 
\\
&\qquad
+ 
T^\psi_{\mu\nu}(f^-_I, f_J^+)
\hat{d}_I \hat{b}_J 
+ 
T^\psi_{\mu\nu}(f^-_I, f^-_J)
\hat{d}_I \hat{d}^\dag_J
\biggr],
\end{split}
\end{equation}
where
\begin{equation} \label{T(fI, fJ)}
\begin{split}
T^\psi_{\mu\nu}(f_I, f_J) 
&\equiv -\frac{1}{4} \biggl[
\bar{f}_I \gamma_\mu D_\nu f_J
+ \bar{f}_I \gamma_\nu D_\mu f_J
\\
&\qquad
- (D_\mu \bar{f}_I) \gamma_\nu f_J
- (D_\nu \bar{f}_I) \gamma_\mu f_J
\biggr].
\end{split}
\end{equation}
Just as with the electromagnetic current, the expectation value of Eq.~(\ref{hat T}) is divergent since Eq.~(\ref{<Tpsi>}) contains products of the field operator.  Again following \cite{Alcubierre:2022rgp, Kain:2023jgu}, we use normal ordering,
\begin{equation} \label{EFE 2}
G_{\mu\nu} = 8\pi G 
\left( \langle {:\mathrel{\widehat{T}^\psi_{\mu\nu}}:} \rangle + T^\mathcal{A}_{\mu\nu} \right),
\end{equation}
where 
\begin{equation} \label{<Tpsi> 2}
\begin{split}
{:\mathrel{\widehat{T}^\psi_{\mu\nu}}:}
&= \sum_{I,J} \biggl[
T^\psi_{\mu\nu}(f_I^+, f_J^+)
\hat{b}^\dag_I \hat{b}_J 
+ 
T^\psi_{\mu\nu}(f_I^+, f^-_J)
\hat{b}^\dag_I \hat{d}^\dag_J 
\\
&\qquad
+ 
T^\psi_{\mu\nu}(f^-_I, f_J^+)
\hat{d}_I \hat{b}_J 
- 
T^\psi_{\mu\nu}(f^-_I, f^-_J)
\hat{d}^\dag_J\hat{d}_I 
\biggr].
\end{split}
\end{equation}

The Einstein field equations in (\ref{EFE 2}) lead to the following equations for the metric fields \cite{Calhoun:2022xrw},
\begin{equation} \label{metric field eom}
\begin{split}
\partial^2_r \alpha& = 
\left(
\frac{\partial_r A}{2A}
- \frac{\partial_r C}{ C}
\right) \partial_r \alpha
+ 4\pi G \alpha A (\rho + S )
\\
\partial_r^2 C &=
A + \frac{(\partial_r A)(\partial_r C)}{2A} 
+ \frac{(\partial_r C)^2}{4C}
- 8\pi G AC \rho
\\
0
&= \frac{1}{C} - \frac{(\partial_r \alpha)(\partial_r C)}{\alpha AC}  
- \frac{(\partial_r C)^2 }{4AC^2} 
+ 8\pi  G S\indices{^r_r},
\end{split}
\end{equation}
where
\begin{equation} \label{matter functions}
\begin{split}
\rho &= \frac{1}{\alpha^2}
\left( \langle {:\mathrel{\widehat{T}^\psi_{tt}}:} \rangle + 
T^\mathcal{A}_{tt} \right)
\\
S\indices{^r_r} &= \frac{1}{A}
\left( \langle {:\mathrel{\widehat{T}^\psi_{rr}}:} \rangle + 
T^\mathcal{A}_{rr} \right)
\\
S\indices{^\theta_\theta} &= \frac{1}{C}
\left( \langle {:\mathrel{\widehat{T}^\psi_{\theta\theta}}:} \rangle +
T^\mathcal{A}_{\theta\theta} \right)
\end{split}
\end{equation}
are the energy density and spatial stress and where $S = S\indices{^r_r} + 2S\indices{^\theta_\theta}$ is the trace of the spatial stress.


\section{Spherically symmetric states and expectation values}
\label{sec:states}

In Sec.~\ref{sec:Dirac}, we constructed creation and annihilation operators for the quantized Dirac field.  These operators satisfy the standard anticommutation relations, as can be see in Eq.~(\ref{b d anticommutation}).  Using these operators we can construct basis states for the Hilbert space,
\begin{equation} \label{basis state}
\begin{split}
&|N_{I_1}^b, N_{I_2}^b, \ldots,N_{J_1}^d, N_{J_2}^d,\ldots\rangle
\\
&\qquad = 
\cdots (\hat{d}^\dag_{J_2})^{N_{J_2}^d} (\hat{d}^\dag_{J_1})^{N_{J_1}^d}
\cdots (\hat{b}^\dag_{I_2})^{N_{I_2}^b} (\hat{b}^\dag_{I_1})^{N_{I_1}^b} 
|0\rangle,
\end{split}
\end{equation} 
where $N_I^b$ and $N_J^d$ are occupation numbers and $|0\rangle$ is the vacuum state.  The vacuum state is normalized and annihilated by the annihilation operators, $\hat{b}_I |0\rangle = \hat{d}_J |0\rangle = 0$.  The states $|N_{I_1}^b, N_{I_2}^b, \ldots,N_{J_1}^d, N_{J_2}^d,\ldots\rangle$ form an orthonormal basis for the Hilbert space and are eigenstates of the number operators,
\begin{equation}
\begin{split}
\widehat{N}_I^b |N_{I_1}^b, \ldots,N_{J_1}^d,\ldots\rangle
&= N_I^b |N_{I_1}^b, \ldots,N_{J_1}^d,\ldots\rangle
\\
\widehat{N}_J^d |N_{I_1}^b, \ldots,N_{J_1}^d,\ldots\rangle
&= N_J^d |N_{I_1}^b, \ldots,N_{J_1}^d,\ldots\rangle,
\end{split}
\end{equation}
where 
\begin{equation}
\widehat{N}_I^b = \hat{b}^\dag_I \hat{b}_I, \qquad
\widehat{N}_J^d = \hat{d}^\dag_J \hat{d}_J.
\end{equation}
Being fermionic states, the occupation numbers can have values $N^b_I, N_J^d=0 \text{ or } 1$.

In Sec.~\ref{sec:Maxwell}, we used a semiclassical approximation and wrote the gauge field equations of motion in Eq.~(\ref{gauge eom 2}) in terms of the expectation value of the electromagnetic current.  In Sec.~\ref{sec:Einstein}, we again used a semiclassical approximation and wrote the Einstein field equations in Eq.~(\ref{EFE 2}) in terms of the expectation value of the stress-energy-momentum tensor for the Dirac field and the classical stress-energy-momentum tensor for the gauge field.  We choose to evaluate the expectation values using the basis states in (\ref{basis state}), for which
\begin{equation} 
\begin{split}
\langle{:\mathrel{\hat{j}^\nu}:}\rangle
&= \sum_I 
\biggl[ 
N_I^b
j^\nu(f_I^+, f_I^+)
- 
N_I^d
j^\nu(f_I^-, f_I^-)
\biggr]
\\
\langle{:\mathrel{\widehat{T}^\psi_{\mu\nu}}:}\rangle
&= \sum_I \biggl[
N_I^b
T_{\mu\nu}(f_I^+, f_I^+)
- 
N_I^d
T_{\mu\nu}(f^-_I, f^-_I)
\biggr].
\end{split}
\end{equation}

Since we have a spherically symmetric spacetime, we must use spherically symmetric basis states.  Spherically symmetric basis states have zero total angular momentum.  As shown in \cite{Finster:1998ju, Olabarrieta:2007di, Alcubierre:2022rgp, Kain:2023jgu}, zero total angular momentum is achieved by exciting all possible values of $m_j$ for each excited $j$.  The expectation values we make use of are then
\begin{equation} \label{<T>}
\begin{split}
\langle {:\mathrel{\hat{j}^\nu}:} \rangle 
&= \sum_{n,j} N_{nj}^b \sum_{m_j = -j}^j
j^\nu(f_{njm_j+}, f_{njm_j+})
\\
&\qquad
- \sum_{n,j} N_{nj}^d \sum_{m_j = -j}^j
j^\nu(f_{njm_j-}, f_{njm_j-})
\\
\langle {:\mathrel{\widehat{T}^\psi_{\mu\nu}}:} \rangle 
&= \sum_{n,j} N_{nj}^b \sum_{m_j = -j}^j
T^\psi_{\mu\nu}(f_{njm_j+}, f_{njm_j+})
\\
&\qquad
- \sum_{n,j} N_{nj}^d \sum_{m_j = -j}^j
T^\psi_{\mu\nu}(f_{njm_j-}, f_{njm_j-}).
\end{split}
\end{equation}

From the gauge field equations of motion in (\ref{gauge field eom}), we can see that the only component of the expectation value of the electromagnetic current that we need is $\langle {:\mathrel{\hat{j}^t}:} \rangle$.  In a static spherically symmetric spacetime, the stress-energy-momentum tensor must be diagonal.  Details for how some of these quantities are computed are given in Appendix B of \cite{Kain:2023jgu}.  For the electromagnetic current we find
\begin{equation} \label{j(fI,fJ)}
\sum_{m_j=-j}^j j^t (f_I^\pm, f_I^\pm)
= - \frac{e(2j+1)}{4\pi \alpha^2 C}
\left(P_{nj+}^2 + P_{nj-}^2 \right),
\end{equation}
for the stress-energy-momentum tensor for the Dirac field we find
\begin{align} 
\sum_{m_j=-j}^j T^\psi_{tt}(f_I^\pm, f_I^\pm)
&= \pm (e\mathcal{A}_t + \omega_{nj})
\frac{2j+1}{4\pi C} 
\left(P_{nj+}^2 + P_{nj-}^2 \right)
\notag
\\
\sum_{m_j=-j}^j T^\psi_{rr}(f_I^\pm, f_I^\pm)
&= \pm
\frac{A(2j+1)}{4\pi \alpha^2 C}
\biggl[ 
- \frac{\alpha(2j+1)}{\sqrt{C}} 
P_{nj+} P_{nj-}
\notag
\\
&\qquad
+ (e\mathcal{A}_t + \omega_{nj} - \alpha m_\psi) P_{nj+}^2
\notag
\\
&\qquad
+ (e\mathcal{A}_t + \omega_{nj} + \alpha m_\psi) P_{nj-}^2
\biggr]
\notag
\\
\sum_{m_j=-j}^j T^\psi_{\theta\theta}(f_I^\pm, f_I^\pm) 
&= \pm \frac{(2j+1)^2}{8\pi \alpha \sqrt{C}} P_{nj+} P_{nj-}
\notag
\\
\sum_{m_j=-j}^j T^\psi_{\phi\phi}(f_I^\pm, f_I^\pm) 
&= \pm \frac{(2j+1)^2}{8\pi \alpha \sqrt{C}} P_{nj+} P_{nj-} \sin^2\theta,
\label{T(fI,fJ)}
\end{align}
and for the stress-energy-momentum tensor for the gauge field, which follows from the bottom equation in (\ref{SEMT}), we find
\begin{equation} \label{EM tensore gauge}
\begin{split}
T^\mathcal{A}_{tt} 
&= \frac{(\partial_r \mathcal{A}_t)^2}{2 A}
\\
T^\mathcal{A}_{rr} 
&= -\frac{(\partial_r \mathcal{A}_t)^2}{2 \alpha^2}
\\
T^\mathcal{A}_{\theta\theta}
&= \frac{C(\partial_r \mathcal{A}_t)^2}{2 \alpha^2 A}
\\
T^\mathcal{A}_{\phi\phi}
&= \frac{C(\partial_r \mathcal{A}_t)^2}{2 \alpha^2 A} \sin^2\theta.
\end{split}
\end{equation}


\section{Einstein-Dirac-Maxwell wormholes}
\label{sec:EDM wormholes}

EDM wormholes are described by self-consistent solutions to the radial equations of motion for the Dirac field in (\ref{radial eom}), the normalization requirement in (\ref{norm}), the equations of motion for the gauge field in (\ref{gauge field eom}), and the metric field equations in (\ref{metric field eom}).  The gauge field equations of motion depend on the expectation value of the electromagnetic current in Eqs.~(\ref{<T>}) and (\ref{j(fI,fJ)}).  The metric field equations depend on the energy density and spatial stress given in (\ref{matter functions}) which depend on stress-energy-momentum tensor equations in (\ref{<T>}), (\ref{T(fI,fJ)}), and (\ref{EM tensore gauge}).

For simplicity, we restrict our attention to $b$-type particles and set $N_{nj}^d = 0$.  In this case, the quantum numbers that distinguish solutions are $n$ and $j$.  For a single value of $n$ and for $j=1/2$, our system of equations is equivalent to the system of equations in \cite{Konoplya:2021hsm, Kain:2023ann}.  However, these papers maintained spherical symmetry by including two Dirac fields and the fields were not quantized.  In our formalism, we have a single Dirac field and we maintain spherical symmetry by computing expectation values with states formed from spherically symmetric excitations of the quantized Dirac field. Further, our formalism allows us to consider multiple values of $n$ and $j$ and allows for values of $j$ larger than $1/2$.

To numerically solve the system of equations, we need to make a coordinate choice for the metric functions.  For example, a common choice in the study of wormholes is $C(r) = R^2_0 + r^2$, where $R_0$ is the wormhole throat radius defined in Eq.~(\ref{R0 def}).  Another possibility is $C(r) = R_0^2[1-(r/R_0)^2]^{-2}$, which was used in \cite{Konoplya:2021hsm} and which compactifies the radial coordinate to $-R_0 < r < R_0$.  We prefer the coordinate choice \cite{Kain:2023ann}
\begin{equation} \label{A=1}
A(r) = 1.
\end{equation}
This choice sets the radial coordinate equal to the physical distance from the origin.

Just as in \cite{Kain:2023jgu}, we make use of scaling symmetries to write the system of equations in terms of dimensionless variables.  In the first scaling, we use the wormhole throat radius, $R_0$, to define the dimensionless variables
\begin{equation}
\begin{gathered}
\bar{r} \equiv \frac{r}{R_0},
\quad
\bar{\omega}_{nj} \equiv R_0 \omega_{nj},
\quad
\bar{e} \equiv \frac{R_0}{\sqrt{G}} e,
\quad
\bar{m}_\psi \equiv R_0 m_\psi,
\\
\overline{P}_{nj\pm} \equiv \sqrt{\frac{G}{R_0}} P_{nj\pm},
\quad
\bar{\mathcal{A}_t} \equiv \sqrt{G} \mathcal{A}_t,
\quad
\overline{C} \equiv \frac{1}{R_0^2} C.
\end{gathered} 
\end{equation}
In the second scaling, we scale variables by $\alpha(0)$, which helps with specifying boundary conditions, by defining
\begin{equation} \label{alpha scale}
\begin{aligned}
\widetilde{\alpha}(\bar{r}) &\equiv \frac{\alpha(\bar{r})}{\alpha(0)},& 
\qquad
\widetilde{P}_{nj\pm}(\bar{r}) &\equiv \frac{\overline{P}_{nj\pm}(\bar{r})}{\sqrt{\alpha(0)}}
\\
\widetilde{\mathcal{A}}_t(\bar{r}) &\equiv \frac{\bar{\mathcal{A}}_t(\bar{r})}{\alpha(0)},&
\widetilde{\omega}_{nj} &\equiv \frac{\bar{\omega}_{nj}}{\alpha(0)}.
\end{aligned}
\end{equation}
When written in terms of these variables, $G$, $R_0$, and $\alpha(0)$ cancel out.

In terms of the dimensionless variables, Eq.~(\ref{(fI,fI)}) for the norm of a mode function becomes
\begin{equation} 
\mathcal{N}_{nj} 
= \left(\frac{R_0}{\ell_P}\right)^2 \overline{\mathcal{N}}_{nj},
\quad
\overline{\mathcal{N}}_{nj} 
\equiv
\int_{-\infty}^\infty d\bar{r} \frac{1}{\widetilde{\alpha}}
\biggl( 
\widetilde{P}_{nj+}^2
+
\widetilde{P}_{nj-}^2
\biggr),
\end{equation}
where $\ell_P = \sqrt{G}$ is the Planck length and where we used the coordinate choice in (\ref{A=1}).  Imposing the normalization requirement in (\ref{norm}), we have
\begin{equation} \label{R_0 eq}
\frac{R_0}{\ell_P} = \frac{1}{\sqrt{\overline{\mathcal{N}}_{nj}}}.
\end{equation}
We find that all $\overline{\mathcal{N}}_{nj}$ must equal the same value and that solutions are only valid for a single wormhole throat radius, $R_0$, as given by (\ref{R_0 eq}), since only for this value are the mode functions normalized.

To integrate the system of equations outward, we need inner boundary conditions at $\bar{r} = 0$.  The scaled variables obey
\begin{equation}
\overline{C}(0) = 1, \qquad 
\widetilde{\alpha}(0) = 1, \qquad 
\widetilde{\mathcal{A}}_t(0) = 0,
\end{equation}
where the last condition follows from the gauge choice in (\ref{At(0) = 0}).  Additionally, we will look for solutions that satisfy \cite{Konoplya:2021hsm, Kain:2023ann}
\begin{equation}
\overline{C}'(0) = 0, \qquad
\widetilde{\alpha}'(0) = 0,
\end{equation}
where a prime indicates a derivative with respect to $\bar{r}$.  Plugging Taylor series expansions of the fields into the scaled version of the radial equations of motion, we find
\begin{align}
\widetilde{P}_{nj\pm} (0) &=
p_{nj\pm},
\end{align}
where the $p_{nj\pm}$ are undetermined constants.  Doing the same with the scaled version of the bottom equation in (\ref{metric field eom}), we find
\begin{align} \label{At'(0)}
\widetilde{\mathcal{A}}^{\prime \, 2}_t(0)
&= 
\frac{1}{4\pi} \Bigl\{
1 + 
\sum_{n,j} N^b_{nj} 2(2j+1)
\notag
\\
&\qquad 
\times
\bigl[
(\widetilde{\omega}_{nj} - \bar{m}_\psi)p_{nj+}^2 
+ (\widetilde{\omega}_{nj} + \bar{m}_\psi)p_{nj-}^2
\notag
\\
&\qquad
- (2j + 1)p_{nj+}p_{nj-}
\bigr]
\Bigr\}.
\end{align}
Inner boundary conditions are parametrized in terms of the constants $p_{nj+}$ and $p_{nj-}$.  Without loss of generality, we use the positive root in Eq.~(\ref{At'(0)});~using the negative root is equivalent to using the positive root but with the sign of $e$ flipped.

\begin{figure*}
\centering
\includegraphics[width=6.5in]{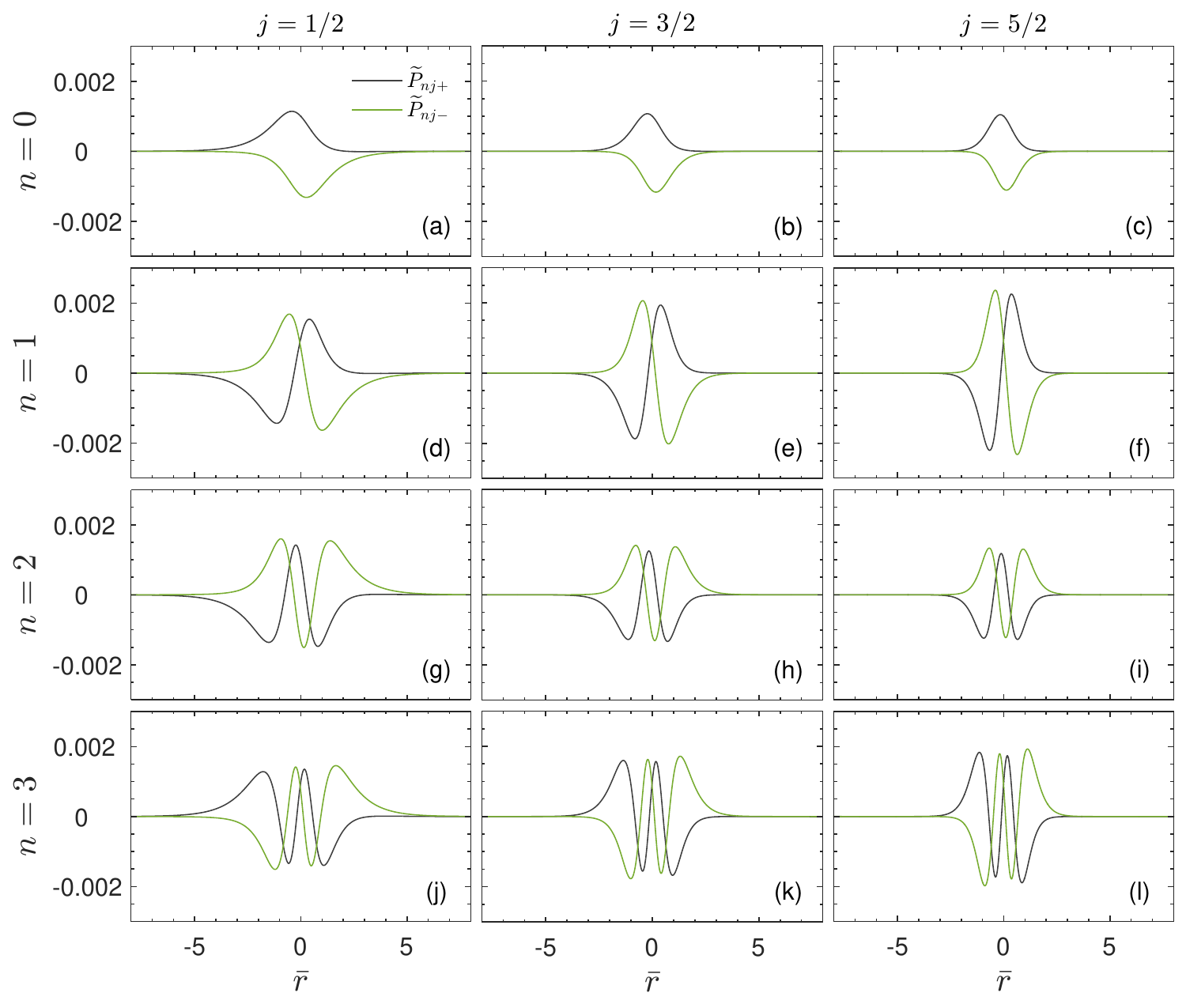}
\caption{Semiclassical EDM wormhole configurations with a single $n$ and a single $j$ excited.  Each figure plots the radial functions $\widetilde{P}_{nj\pm}(\bar{r})$ for $p_{nj+} = 0.001$, $\bar{e} = 0.1$, $\bar{m}_\psi = 0.2$, and $n$ and $j$ as indicated.}
\label{fig:1}
\end{figure*}

\begin{figure*}
\centering
\includegraphics[width=7in]{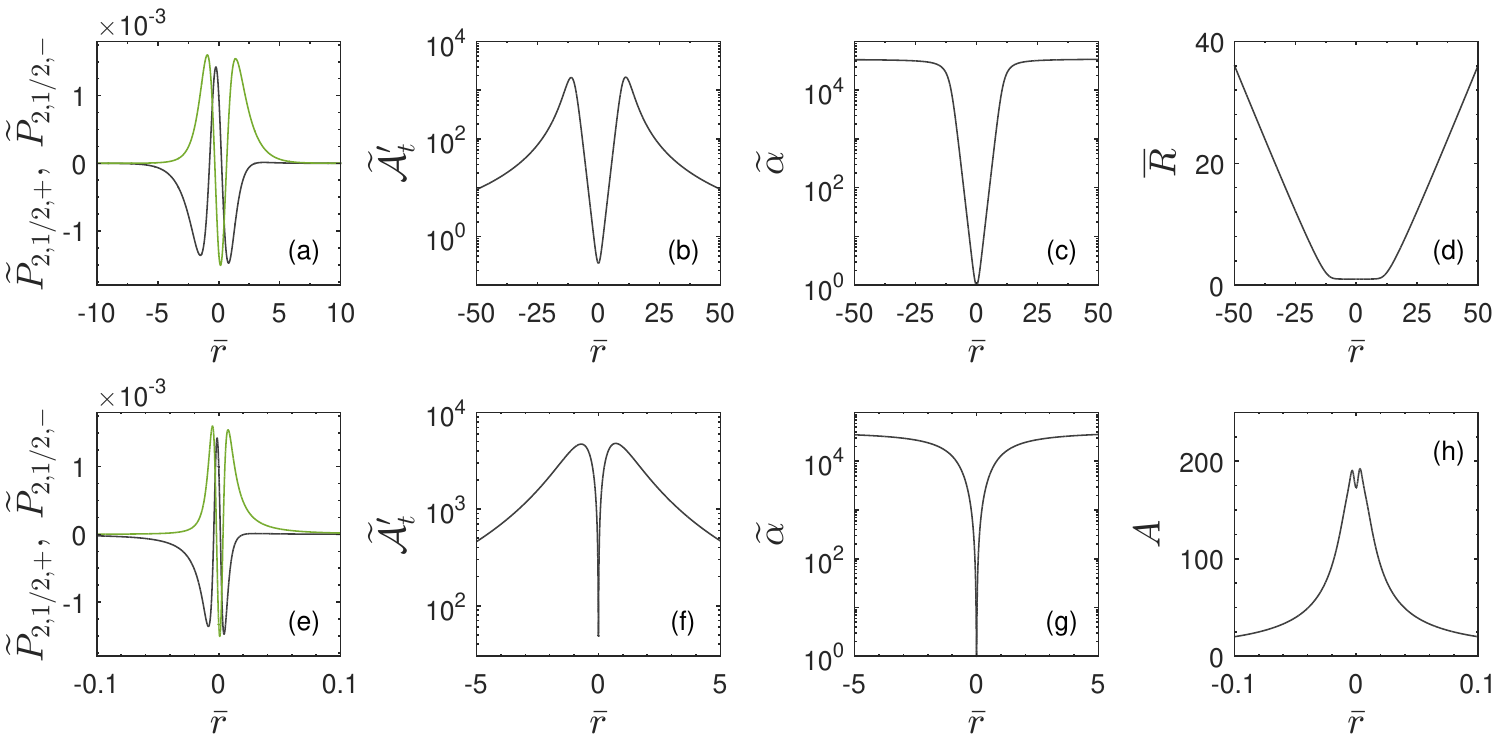}
\caption{(a--d) display fields for the same solution shown in Fig.~\ref{fig:1}(g) with $p_{nj+} = 0.001$, $\bar{e} = 0.1$, $\bar{m}_\psi = 0.2$, $n = 2$, and $j=1/2$.  (e--h) display fields for the same configuration as the top row, except using the coordinate choice $\overline{R}^2(\bar{r}) = \overline{C}(\bar{r}) = 1 + \bar{r}^2$ instead of $A=1$.}
\label{fig:2}
\end{figure*}

\setlength{\tabcolsep}{9pt}
\begin{table} 
\normalsize
\begin{tabular}{ccccc}
$n,\, j$ & $p_{nj+}$ & $p_{nj-}$ &  $\widetilde{\omega}_{nj}$ & $R_0/\ell_P$
\\
\hline\hline
$0,\, 1/2$ & 0.001 & $-0.001255$ & $-1.519$ & 499.0
\\
$0,\, 3/2$ & 0.001 & $-0.001121$ & $-2.510$ & 614.0
\\
$0,\, 5/2$ & 0.001 & $-0.001079$ & $-3.506$ & 686.3
\\
$1,\, 1/2$ & 0.001 & $+0.000709$ & $-2.519$ & 375.3
\\
$1,\, 3/2$ & 0.001 & $+0.000796$ & $-3.510$ & 322.5 
\\
$1,\, 5/2$ & 0.001 & $+0.000827$ & $-4.506$ & 294.2
\\
$2,\, 1/2$ & 0.001 & $-0.001329$ & $-3.519$ & 392.2
\\
$2,\, 3/2$ & 0.001 & $-0.001164$ & $-4.510$ & 465.2
\\
$2,\, 5/2$ & 0.001 & $-0.001110$ & $-5.506$ & 510.6
\\
$3,\, 1/2$ & 0.001 & $+0.000695$ & $-4.519$ & 414.2
\\
$3,\, 3/2$ & 0.001 & $+0.000786$ & $-5.509$ & 367.6
\\
$3,\, 5/2$ & 0.001 & $+0.000819$ & $-6.505$ & 341.0
\end{tabular}
\caption{Parameter values for all example solutions shown in Fig.~\ref{fig:1} with $\bar{e} = 0.1$ and $\bar{m}_\psi = 0.2$.  We note that the $\widetilde{\omega}_{nj}$ are gauge dependent and that the listed values are specific to the gauge choice in Eqs.~(\ref{Ar = 0}) and (\ref{At(0) = 0}).}
\label{table1}
\end{table}

\begin{figure*}
\centering
\includegraphics[width=7in]{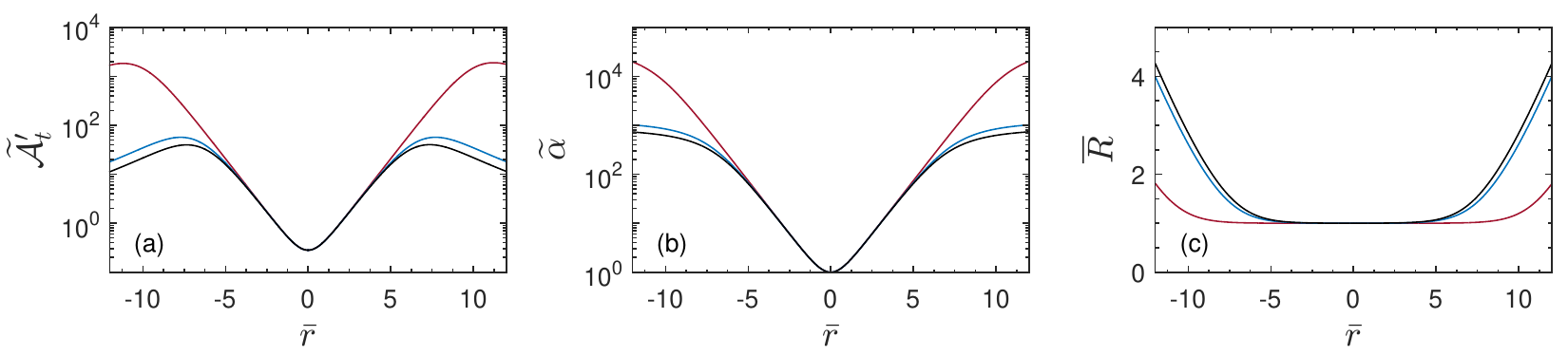}
\caption{$\widetilde{A}_t'$, $\widetilde{\alpha}$, and $\overline{R}$ are plotted for three solutions.  Each solution has $\bar{e} = 0.1$ and $\bar{m}_\psi = 0.2$.  The solutions are then defined by $p_{nj+} = 0.001$, $ n = 0$, $j=5/2$ (red curves);~$p_{nj+} = 0.003$, $ n = 1$, $j=3/2$ (blue curves);~and $p_{nj+} = 0.005$, $ n = 2$, $j=1/2$ (black curves).  Around the origin these fields are nearly identical, suggesting that spacetime and the electromagnetic field may be universal near the wormhole.}
\label{fig:3}
\end{figure*}

\begin{figure}
\centering
\includegraphics[width=2.75in]{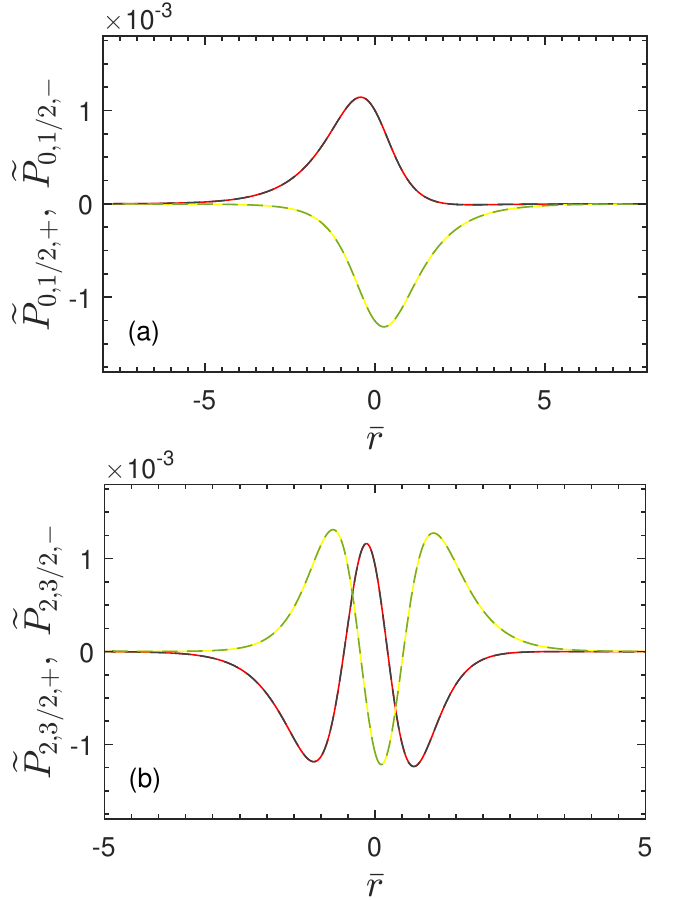}
\caption{Multi-$n$, multi-$j$ semiclassical EDM wormhole configuration with $(n,j) = (0,1/2) \text{ and } (2,1/2)$ excited, with $\bar{e} = 0.1$, $\bar{m}_\psi = 0.2$, and $p_{0,1/2,+} = 0.001$.  The solid black and green curves plot the radial functions for the multi-$n$, multi-$j$ solution.  The dashed red and yellow curves plot the corresponding and independent single-$n$, single-$j$ solutions, which agree very well with the multi-$n$, multi-$j$ solution.}
\label{fig:4}
\end{figure}

\begin{figure}
\centering
\includegraphics[width=2.75in]{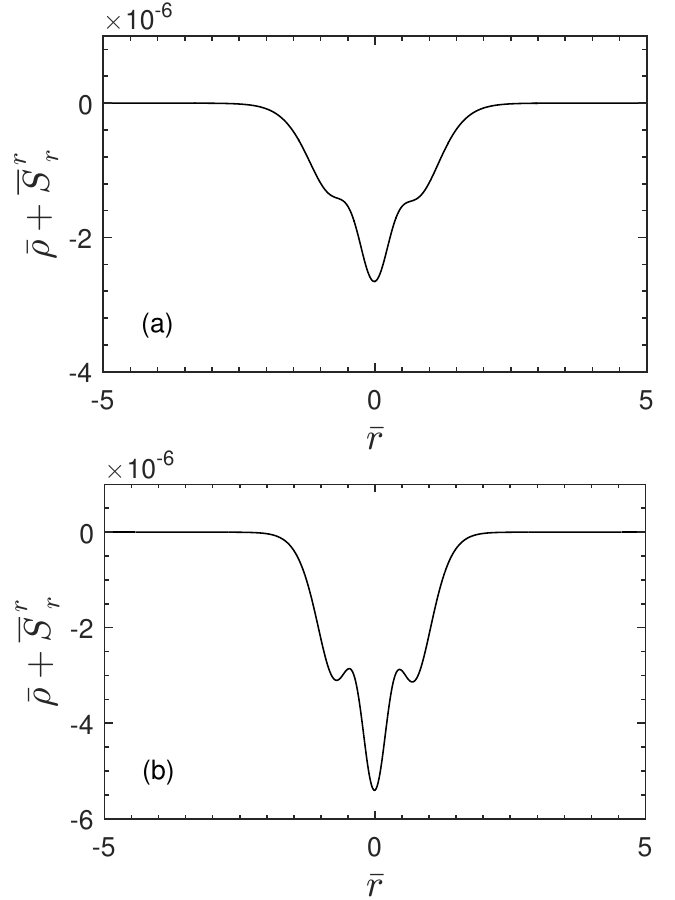}
\caption{The null energy condition is violated if $\rho + S\indices{^r_r} < 0$.  We plot (the dimensionless) $\bar{\rho} + \overline{S}\indices{^r_r}$ in (a) for the wormhole shown in Figs.~\ref{fig:2}(a)--\ref{fig:2}(d) and plot $\bar{\rho} + \overline{S}\indices{^r_r}$ in (b) for the multi-$n$, multi-$j$ wormhole shown in Fig.~\ref{fig:4}.  In both cases we can see that the null energy condition is violated, as expected for a wormhole.}
\label{fig:5}
\end{figure}

For outer boundary conditions, we require that the energy density $\rho$, which is defined in (\ref{matter functions}), asymptotically heads to zero at spatial infinity, i.e.~$\rho \rightarrow 0$ for $\bar{r} \rightarrow \pm \infty$.  This is accomplished by requiring $\widetilde{P}_{nj\pm}, \,  \widetilde{\mathcal{A}}'_t \rightarrow 0$ for $\bar{r} \rightarrow \pm \infty$.

Solutions require specification of the constants $p_{nj\pm}$, $\widetilde{\omega}_{nj}$, $\bar{e}$, and $\bar{m}_\psi$.  Further, the solutions must satisfy the normalization requirement in (\ref{R_0 eq}), which in many cases is a nontrivial constraint.


\subsection{Single-$n$, single-$j$ examples}
\label{sec:single n j}

Consider a wormhole that has a single value of $n$ and a single value of $j$ excited.  In this case, there is a single $\overline{\mathcal{N}}_{nj}$ and the normalization constraint is trivially satisfied by any solution.  To find a solution, we must determine $p_{nj+}$, $p_{nj-}$, and $\widetilde{\omega}_{nj}$, along with $\bar{e}$ and $\bar{m}_\psi$.  One strategy is to specify $p_{nj+}$, $\bar{e}$, and $\bar{m}_\psi$ and to use trial values for $p_{nj-}$ and $\widetilde{\omega}_{nj}$.  Using the shooting method, we can integrate the system of equations outward from $\bar{r} = 0$ and tune $p_{nj-}$ and $\widetilde{\omega}_{nj}$ until the outer boundary conditions are satisfied.  In this case, we have two shooting parameters.  We used this method to find static solutions in \cite{Kain:2023ann}.  In Fig.~\ref{fig:1}, we show example solutions for the radial functions $\widetilde{P}_{nj\pm}(\bar{r})$ for a single $n$ and a single $j$ excited.  We can see that the quantum number $n$ counts the number of radial nodes.  Values for various parameters for these example solutions are listed in Table \ref{table1}.  

In the top row of Fig.~\ref{fig:2}, we show additional fields for the same solution shown in Fig.~\ref{fig:1}(g) with $n=2$ and $j=1/2$.  For comparison, in the bottom row of Fig.~\ref{fig:2} we show the same configuration with $n=2$ and $j=1/2$, but using the coordinate choice $C(r) = R_0^2 + r^2$ instead of $A=1$.  Notice that, in both rows of Fig.~\ref{fig:2}, $\widetilde{\alpha}$ reaches relatively large values as $\bar{r}\rightarrow \pm\infty$.  As a consequence, if we would like the metric to take the standard asymptotically flat form on, say, the positive side such that $\alpha(r\rightarrow +\infty) \rightarrow 1$, then from (\ref{alpha scale}) $\alpha(0) = 1/ \widetilde{\alpha}(\bar{r}\rightarrow +\infty)$ and
\begin{equation}
\alpha(\bar{r}) = \frac{\widetilde{\alpha}(\bar{r})
}{\widetilde{\alpha}(\bar{r}\rightarrow +\infty)}.
\end{equation}
This tells us that $\alpha$ is effectively zero around the origin.  Notice also in Fig.~\ref{fig:2}(h) that the metric function $A$ has a relatively large peak near the origin.

The collapse of $\alpha$ near the origin and the large peak of $A$ strongly suggest that the static configuration shown in Fig.~\ref{fig:2} wants to collapse to a black hole.  Indeed, in \cite{Kain:2023ann} we used $n=0, \, j=1/2$ and $n=1, \, j=1/2$ static solutions as initial data and numerically evolved them forward in time and showed that they collapse.  More specifically, we found that black holes form on both sides of the wormhole such that the black holes are connected by the wormhole.  We speculate that all static solutions considered in the current work would collapse similarly.

In Fig.~\ref{fig:3}, we display fields that are not the radial functions for a few different solutions.  Notice that these fields are nearly identical around the origin, but differ at large $\bar{r}$.  On the other hand, the radial functions differ significantly around the origin, as can be seen in Fig.~\ref{fig:1}.  Figure \ref{fig:3} suggests that spacetime and the electromagnetic field are nearly universal in the vicinity of the wormhole, in that they are effectively independent of the parameters listed in Table \ref{table1} and the radial functions.  In addition to being interesting in its own right, this observation has beneficial consequences for finding multi-$n$, multi-$j$ solutions, as we explain in the next subsection.


\subsection{Multi-$n$, multi-$j$ example}

We now consider wormholes with two pairs of $(n,j)$ excited.  In this case, solutions require specification of two $p_{nj+}$, two $p_{nj-}$, and two $\widetilde{\omega}_{nj}$, along with $\bar{e}$ and $\bar{m}_\psi$.  Further, the normalization requirement is no longer trivially satisfied and becomes a nontrivial constraint on solutions.  If we specify $\bar{e}$, $\bar{m}_\psi$, and one of the $p_{nj+}$, we still have five parameters to determine.  With five shooting parameters, it is time consuming for the shooting method to converge to a solution.  However, we found in the previous subsection that spacetime and the electromagnetic field are nearly universal in the vicinity of the wormhole.  Consequently, the radial functions for the multi-$n$, multi-$j$ solution should be roughly the same as the radial functions for the single-$n$, single-$j$ solutions.  

In practice, we take the initial trial values for the five shooting parameters to be the parameter values for the single-$n$, single-$j$ solutions.  For example, consider the $n = 0, \, j = 1/2, \, p_{nj+} = 0.001$ solution listed in Table \ref{table1}.  This solution has $R_0/\ell_p = 499.0$.  If we would like the second excitation to have, say, $n=2, \, j = 3/2$, we first search for a single-$n$, single-$j$ solution with $n=2, \, j = 3/2, R_0/\ell_p = 499.0$.  We then use the parameter values for these two single-$n$, single-$j$ solutions as the initial trial values in the five parameter shooting method.  We still tune the five shooting parameters so as to find the precise multi-$n$, multi-$j$ solution and to confirm that it is in fact a solution, but the shooting method converges significantly faster than if we had used arbitrary initial trial values.  We show the radial functions for this example in Fig.~\ref{fig:4}.  The solid black and green curves are for the multi-$n$, multi-$j$ solution.  The dashed red and yellow curves are for the two independent single-$n$, single-$j$ solutions.  We can see that they match well, as expected.


\subsection{Null energy condition}

Last, we consider the null energy condition, which must be violated for a wormhole to be open \cite{Morris:1988cz}.  For radial null vectors $d^\mu = d^t(1, \pm \alpha/\sqrt{A},0,0)$, where $d^t$ is an arbitrary constant, the null energy condition is violated if 
\begin{equation}
\langle {:\mathrel{\widehat{T}_{\mu\nu}}:} \rangle d^\mu d^\nu < 0,
\end{equation}
which is equivalent to
\begin{equation}
\rho + S\indices{^r_r} < 0.
\end{equation}
We display $\rho + S\indices{^r_r}$ in Fig.~\ref{fig:5} for the same wormholes shown in Figs.~\ref{fig:2} and \ref{fig:4}.  As can be seen, $\rho + S\indices{^r_r} < 0$ and these wormholes violate the null energy condition, as expected.

It is sometimes stated that a wormhole is traversable if the null energy condition is violated.  However, we showed in \cite{Kain:2023ann} that $n=0,\, j=1/2$ and $n=1,\,j=1/2$ wormholes collapse sufficiently quickly that any null geodesic that travels through the wormhole will be caught inside a black hole and will not be able to travel arbitrarily far on the opposite side of the wormhole.  Violation of the null energy condition is sufficient for the wormhole to be open, but for EDM wormholes it is insufficient for the wormhole to be traversable.


\section{Conclusion}
\label{sec:conclusion}

EDM wormholes are perhaps the only wormhole solutions discovered that do not make use of exotic matter and exist in asymptotically flat general relativity.  Previous constructions of EDM wormholes used multiple independent Dirac fields and treated gravity and the electromagnetic field classically.  In this work, we used a single Dirac field and constructed EDM wormhole configurations in quantum field theory with gravity and the electromagnetic field treated semiclassically.  Our framework puts EDM wormholes on a more secure theoretical footing and is able to describe a broader class of wormhole configurations than previously considered.  In particular, our framework can describe configurations with total angular momentum $j > 1/2$ and multi-$n$, multi-$j$ configurations, examples of which we presented.

We also showed that spacetime and the electromagnetic field may have a universal structure in the vicinity of the wormhole, even though the Dirac field does not.  We speculated that all configurations considered would collapse and form black holes that are connected by the wormhole, analogously to the collapse studied in \cite{Kain:2023ann}.




\begin{thebibliography}{29}%
\makeatletter
\providecommand \@ifxundefined [1]{%
 \@ifx{#1\undefined}
}%
\providecommand \@ifnum [1]{%
 \ifnum #1\expandafter \@firstoftwo
 \else \expandafter \@secondoftwo
 \fi
}%
\providecommand \@ifx [1]{%
 \ifx #1\expandafter \@firstoftwo
 \else \expandafter \@secondoftwo
 \fi
}%
\providecommand \natexlab [1]{#1}%
\providecommand \enquote  [1]{``#1''}%
\providecommand \bibnamefont  [1]{#1}%
\providecommand \bibfnamefont [1]{#1}%
\providecommand \citenamefont [1]{#1}%
\providecommand \href@noop [0]{\@secondoftwo}%
\providecommand \href [0]{\begingroup \@sanitize@url \@href}%
\providecommand \@href[1]{\@@startlink{#1}\@@href}%
\providecommand \@@href[1]{\endgroup#1\@@endlink}%
\providecommand \@sanitize@url [0]{\catcode `\\12\catcode `\$12\catcode
  `\&12\catcode `\#12\catcode `\^12\catcode `\_12\catcode `\%12\relax}%
\providecommand \@@startlink[1]{}%
\providecommand \@@endlink[0]{}%
\providecommand \url  [0]{\begingroup\@sanitize@url \@url }%
\providecommand \@url [1]{\endgroup\@href {#1}{\urlprefix }}%
\providecommand \urlprefix  [0]{URL }%
\providecommand \Eprint [0]{\href }%
\providecommand \doibase [0]{https://doi.org/}%
\providecommand \selectlanguage [0]{\@gobble}%
\providecommand \bibinfo  [0]{\@secondoftwo}%
\providecommand \bibfield  [0]{\@secondoftwo}%
\providecommand \translation [1]{[#1]}%
\providecommand \BibitemOpen [0]{}%
\providecommand \bibitemStop [0]{}%
\providecommand \bibitemNoStop [0]{.\EOS\space}%
\providecommand \EOS [0]{\spacefactor3000\relax}%
\providecommand \BibitemShut  [1]{\csname bibitem#1\endcsname}%
\let\auto@bib@innerbib\@empty
\bibitem [{\citenamefont {Bl\'azquez-Salcedo}\ \emph
  {et~al.}(2021)\citenamefont {Bl\'azquez-Salcedo}, \citenamefont {Knoll},\
  and\ \citenamefont {Radu}}]{Blazquez-Salcedo:2020czn}%
  \BibitemOpen
  \bibfield  {author} {\bibinfo {author} {\bibfnamefont {J.~L.}\ \bibnamefont
  {Bl\'azquez-Salcedo}}, \bibinfo {author} {\bibfnamefont {C.}~\bibnamefont
  {Knoll}},\ and\ \bibinfo {author} {\bibfnamefont {E.}~\bibnamefont {Radu}},\
  }\bibfield  {title} {\bibinfo {title} {{Traversable wormholes in
  Einstein-Dirac-Maxwell theory}},\ }\href
  {https://doi.org/10.1103/PhysRevLett.126.101102} {\bibfield  {journal}
  {\bibinfo  {journal} {Phys. Rev. Lett.}\ }\textbf {\bibinfo {volume} {126}},\
  \bibinfo {pages} {101102} (\bibinfo {year} {2021})},\ \Eprint
  {https://arxiv.org/abs/2010.07317} {arXiv:2010.07317 [gr-qc]} \BibitemShut
  {NoStop}%
\bibitem [{\citenamefont {Bl\'azquez-Salcedo}\ \emph
  {et~al.}(2022)\citenamefont {Bl\'azquez-Salcedo}, \citenamefont {Knoll},\
  and\ \citenamefont {Radu}}]{Blazquez-Salcedo:2021udn}%
  \BibitemOpen
  \bibfield  {author} {\bibinfo {author} {\bibfnamefont {J.~L.}\ \bibnamefont
  {Bl\'azquez-Salcedo}}, \bibinfo {author} {\bibfnamefont {C.}~\bibnamefont
  {Knoll}},\ and\ \bibinfo {author} {\bibfnamefont {E.}~\bibnamefont {Radu}},\
  }\bibfield  {title} {\bibinfo {title}
  {{Einstein\textendash{}Dirac\textendash{}Maxwell wormholes: ansatz,
  construction and properties of symmetric solutions}},\ }\href
  {https://doi.org/10.1140/epjc/s10052-022-10488-6} {\bibfield  {journal}
  {\bibinfo  {journal} {Eur. Phys. J. C}\ }\textbf {\bibinfo {volume} {82}},\
  \bibinfo {pages} {533} (\bibinfo {year} {2022})},\ \Eprint
  {https://arxiv.org/abs/2108.12187} {arXiv:2108.12187 [gr-qc]} \BibitemShut
  {NoStop}%
\bibitem [{\citenamefont {Konoplya}\ and\ \citenamefont
  {Zhidenko}(2022)}]{Konoplya:2021hsm}%
  \BibitemOpen
  \bibfield  {author} {\bibinfo {author} {\bibfnamefont {R.~A.}\ \bibnamefont
  {Konoplya}}\ and\ \bibinfo {author} {\bibfnamefont {A.}~\bibnamefont
  {Zhidenko}},\ }\bibfield  {title} {\bibinfo {title} {{Traversable Wormholes
  in General Relativity}},\ }\href
  {https://doi.org/10.1103/PhysRevLett.128.091104} {\bibfield  {journal}
  {\bibinfo  {journal} {Phys. Rev. Lett.}\ }\textbf {\bibinfo {volume} {128}},\
  \bibinfo {pages} {091104} (\bibinfo {year} {2022})},\ \Eprint
  {https://arxiv.org/abs/2106.05034} {arXiv:2106.05034 [gr-qc]} \BibitemShut
  {NoStop}%
\bibitem [{\citenamefont {Danielson}\ \emph {et~al.}(2021)\citenamefont
  {Danielson}, \citenamefont {Satishchandran}, \citenamefont {Wald},\ and\
  \citenamefont {Weinbaum}}]{Danielson:2021aor}%
  \BibitemOpen
  \bibfield  {author} {\bibinfo {author} {\bibfnamefont {D.~L.}\ \bibnamefont
  {Danielson}}, \bibinfo {author} {\bibfnamefont {G.}~\bibnamefont
  {Satishchandran}}, \bibinfo {author} {\bibfnamefont {R.~M.}\ \bibnamefont
  {Wald}},\ and\ \bibinfo {author} {\bibfnamefont {R.~J.}\ \bibnamefont
  {Weinbaum}},\ }\bibfield  {title} {\bibinfo {title}
  {{Bl\'azquez-Salcedo\textendash{}Knoll\textendash{}Radu wormholes are not
  solutions to the Einstein-Dirac-Maxwell equations}},\ }\href
  {https://doi.org/10.1103/PhysRevD.104.124055} {\bibfield  {journal} {\bibinfo
   {journal} {Phys. Rev. D}\ }\textbf {\bibinfo {volume} {104}},\ \bibinfo
  {pages} {124055} (\bibinfo {year} {2021})},\ \Eprint
  {https://arxiv.org/abs/2108.13361} {arXiv:2108.13361 [gr-qc]} \BibitemShut
  {NoStop}%
\bibitem [{\citenamefont {Kain}(2023{\natexlab{a}})}]{Kain:2023ann}%
  \BibitemOpen
  \bibfield  {author} {\bibinfo {author} {\bibfnamefont {B.}~\bibnamefont
  {Kain}},\ }\bibfield  {title} {\bibinfo {title} {{Are Einstein-Dirac-Maxwell
  wormholes traversable?}},\ }\href
  {https://doi.org/10.1103/PhysRevD.108.044019} {\bibfield  {journal} {\bibinfo
   {journal} {Phys. Rev. D}\ }\textbf {\bibinfo {volume} {108}},\ \bibinfo
  {pages} {044019} (\bibinfo {year} {2023}{\natexlab{a}})},\ \Eprint
  {https://arxiv.org/abs/2305.11217} {arXiv:2305.11217 [gr-qc]} \BibitemShut
  {NoStop}%
\bibitem [{\citenamefont {Kain}(2023{\natexlab{b}})}]{kain_EREPR}%
  \BibitemOpen
  \bibfield  {author} {\bibinfo {author} {\bibfnamefont {B.}~\bibnamefont
  {Kain}},\ }\bibfield  {title} {\bibinfo {title} {{Probing the Connection
  between Entangled Particles and Wormholes in General Relativity}},\ }\href
  {https://doi.org/10.1103/PhysRevLett.131.101001} {\bibfield  {journal}
  {\bibinfo  {journal} {Phys. Rev. Lett.}\ }\textbf {\bibinfo {volume} {131}},\
  \bibinfo {pages} {101001} (\bibinfo {year} {2023}{\natexlab{b}})},\ \Eprint
  {https://arxiv.org/abs/2309.03314} {arXiv:2309.03314 [hep-th]} \BibitemShut
  {NoStop}%
\bibitem [{\citenamefont {Maldacena}\ and\ \citenamefont
  {Susskind}(2013)}]{Maldacena:2013xja}%
  \BibitemOpen
  \bibfield  {author} {\bibinfo {author} {\bibfnamefont {J.}~\bibnamefont
  {Maldacena}}\ and\ \bibinfo {author} {\bibfnamefont {L.}~\bibnamefont
  {Susskind}},\ }\bibfield  {title} {\bibinfo {title} {{Cool horizons for
  entangled black holes}},\ }\href {https://doi.org/10.1002/prop.201300020}
  {\bibfield  {journal} {\bibinfo  {journal} {Fortsch. Phys.}\ }\textbf
  {\bibinfo {volume} {61}},\ \bibinfo {pages} {781} (\bibinfo {year} {2013})},\
  \Eprint {https://arxiv.org/abs/1306.0533} {arXiv:1306.0533 [hep-th]}
  \BibitemShut {NoStop}%
\bibitem [{\citenamefont {Bolokhov}\ \emph {et~al.}(2021)\citenamefont
  {Bolokhov}, \citenamefont {Bronnikov}, \citenamefont {Krasnikov},\ and\
  \citenamefont {Skvortsova}}]{Bolokhov:2021fil}%
  \BibitemOpen
  \bibfield  {author} {\bibinfo {author} {\bibfnamefont {S.}~\bibnamefont
  {Bolokhov}}, \bibinfo {author} {\bibfnamefont {K.}~\bibnamefont {Bronnikov}},
  \bibinfo {author} {\bibfnamefont {S.}~\bibnamefont {Krasnikov}},\ and\
  \bibinfo {author} {\bibfnamefont {M.}~\bibnamefont {Skvortsova}},\ }\bibfield
   {title} {\bibinfo {title} {{A Note on \textquotedblleft{}Traversable
  Wormholes in Einstein\textendash{}Dirac\textendash{}Maxwell
  Theory\textquotedblright{}}},\ }\href
  {https://doi.org/10.1134/S0202289321040034} {\bibfield  {journal} {\bibinfo
  {journal} {Grav. Cosmol.}\ }\textbf {\bibinfo {volume} {27}},\ \bibinfo
  {pages} {401} (\bibinfo {year} {2021})},\ \Eprint
  {https://arxiv.org/abs/2104.10933} {arXiv:2104.10933 [gr-qc]} \BibitemShut
  {NoStop}%
\bibitem [{\citenamefont {Stuchl\'\i{}k}\ and\ \citenamefont
  {Vrba}(2021)}]{Stuchlik:2021guq}%
  \BibitemOpen
  \bibfield  {author} {\bibinfo {author} {\bibfnamefont {Z.}~\bibnamefont
  {Stuchl\'\i{}k}}\ and\ \bibinfo {author} {\bibfnamefont {J.}~\bibnamefont
  {Vrba}},\ }\bibfield  {title} {\bibinfo {title} {{Epicyclic orbits in the
  field of Einstein\textendash{}Dirac\textendash{}Maxwell traversable wormholes
  applied to the quasiperiodic oscillations observed in microquasars and active
  galactic nuclei}},\ }\href {https://doi.org/10.1140/epjp/s13360-021-02078-4}
  {\bibfield  {journal} {\bibinfo  {journal} {Eur. Phys. J. Plus}\ }\textbf
  {\bibinfo {volume} {136}},\ \bibinfo {pages} {1127} (\bibinfo {year}
  {2021})},\ \Eprint {https://arxiv.org/abs/2110.10569} {arXiv:2110.10569
  [gr-qc]} \BibitemShut {NoStop}%
\bibitem [{\citenamefont {Churilova}\ \emph {et~al.}(2021)\citenamefont
  {Churilova}, \citenamefont {Konoplya}, \citenamefont {Stuchlik},\ and\
  \citenamefont {Zhidenko}}]{Churilova:2021tgn}%
  \BibitemOpen
  \bibfield  {author} {\bibinfo {author} {\bibfnamefont {M.~S.}\ \bibnamefont
  {Churilova}}, \bibinfo {author} {\bibfnamefont {R.~A.}\ \bibnamefont
  {Konoplya}}, \bibinfo {author} {\bibfnamefont {Z.}~\bibnamefont {Stuchlik}},\
  and\ \bibinfo {author} {\bibfnamefont {A.}~\bibnamefont {Zhidenko}},\
  }\bibfield  {title} {\bibinfo {title} {{Wormholes without exotic matter:
  quasinormal modes, echoes and shadows}},\ }\href
  {https://doi.org/10.1088/1475-7516/2021/10/010} {\bibfield  {journal}
  {\bibinfo  {journal} {JCAP}\ }\textbf {\bibinfo {volume} {10}},\ \bibinfo
  {pages} {010}},\ \Eprint {https://arxiv.org/abs/2107.05977} {arXiv:2107.05977
  [gr-qc]} \BibitemShut {NoStop}%
\bibitem [{\citenamefont {Wang}\ \emph {et~al.}(2022)\citenamefont {Wang},
  \citenamefont {Wei},\ and\ \citenamefont {Liu}}]{Wang:2022aze}%
  \BibitemOpen
  \bibfield  {author} {\bibinfo {author} {\bibfnamefont {Y.-Q.}\ \bibnamefont
  {Wang}}, \bibinfo {author} {\bibfnamefont {S.-W.}\ \bibnamefont {Wei}},\ and\
  \bibinfo {author} {\bibfnamefont {Y.-X.}\ \bibnamefont {Liu}},\ }\bibfield
  {title} {\bibinfo {title} {{Comment on ``Traversable Wormholes in General
  Relativity''}},\ }\href@noop {} {\  (\bibinfo {year} {2022})},\ \Eprint
  {https://arxiv.org/abs/2206.12250} {arXiv:2206.12250 [gr-qc]} \BibitemShut
  {NoStop}%
\bibitem [{\citenamefont {Finster}\ \emph
  {et~al.}(1999{\natexlab{a}})\citenamefont {Finster}, \citenamefont
  {Smoller},\ and\ \citenamefont {Yau}}]{Finster:1998ws}%
  \BibitemOpen
  \bibfield  {author} {\bibinfo {author} {\bibfnamefont {F.}~\bibnamefont
  {Finster}}, \bibinfo {author} {\bibfnamefont {J.}~\bibnamefont {Smoller}},\
  and\ \bibinfo {author} {\bibfnamefont {S.-T.}\ \bibnamefont {Yau}},\
  }\bibfield  {title} {\bibinfo {title} {{Particle - like solutions of the
  Einstein-Dirac equations}},\ }\href
  {https://doi.org/10.1103/PhysRevD.59.104020} {\bibfield  {journal} {\bibinfo
  {journal} {Phys. Rev. D}\ }\textbf {\bibinfo {volume} {59}},\ \bibinfo
  {pages} {104020} (\bibinfo {year} {1999}{\natexlab{a}})},\ \Eprint
  {https://arxiv.org/abs/gr-qc/9801079} {arXiv:gr-qc/9801079 [gr-qc]}
  \BibitemShut {NoStop}%
\bibitem [{\citenamefont {Finster}\ \emph
  {et~al.}(1999{\natexlab{b}})\citenamefont {Finster}, \citenamefont
  {Smoller},\ and\ \citenamefont {Yau}}]{Finster:1998ux}%
  \BibitemOpen
  \bibfield  {author} {\bibinfo {author} {\bibfnamefont {F.}~\bibnamefont
  {Finster}}, \bibinfo {author} {\bibfnamefont {J.}~\bibnamefont {Smoller}},\
  and\ \bibinfo {author} {\bibfnamefont {S.-T.}\ \bibnamefont {Yau}},\
  }\bibfield  {title} {\bibinfo {title} {{Particle - like solutions of the
  Einstein-Dirac-Maxwell equations}},\ }\href
  {https://doi.org/10.1016/S0375-9601(99)00457-0} {\bibfield  {journal}
  {\bibinfo  {journal} {Phys. Lett. A}\ }\textbf {\bibinfo {volume} {259}},\
  \bibinfo {pages} {431} (\bibinfo {year} {1999}{\natexlab{b}})},\ \Eprint
  {https://arxiv.org/abs/gr-qc/9802012} {arXiv:gr-qc/9802012 [gr-qc]}
  \BibitemShut {NoStop}%
\bibitem [{\citenamefont {Herdeiro}\ \emph {et~al.}(2017)\citenamefont
  {Herdeiro}, \citenamefont {Pombo},\ and\ \citenamefont
  {Radu}}]{Herdeiro:2017fhv}%
  \BibitemOpen
  \bibfield  {author} {\bibinfo {author} {\bibfnamefont {C.~A.~R.}\
  \bibnamefont {Herdeiro}}, \bibinfo {author} {\bibfnamefont {A.~M.}\
  \bibnamefont {Pombo}},\ and\ \bibinfo {author} {\bibfnamefont
  {E.}~\bibnamefont {Radu}},\ }\bibfield  {title} {\bibinfo {title}
  {{Asymptotically flat scalar, Dirac and Proca stars: discrete vs. continuous
  families of solutions}},\ }\href
  {https://doi.org/10.1016/j.physletb.2017.09.036} {\bibfield  {journal}
  {\bibinfo  {journal} {Phys. Lett. B}\ }\textbf {\bibinfo {volume} {773}},\
  \bibinfo {pages} {654} (\bibinfo {year} {2017})},\ \Eprint
  {https://arxiv.org/abs/1708.05674} {arXiv:1708.05674 [gr-qc]} \BibitemShut
  {NoStop}%
\bibitem [{\citenamefont {Dzhunushaliev}\ and\ \citenamefont
  {Folomeev}(2019)}]{Dzhunushaliev:2018jhj}%
  \BibitemOpen
  \bibfield  {author} {\bibinfo {author} {\bibfnamefont {V.}~\bibnamefont
  {Dzhunushaliev}}\ and\ \bibinfo {author} {\bibfnamefont {V.}~\bibnamefont
  {Folomeev}},\ }\bibfield  {title} {\bibinfo {title} {{Dirac stars supported
  by nonlinear spinor fields}},\ }\href
  {https://doi.org/10.1103/PhysRevD.99.084030} {\bibfield  {journal} {\bibinfo
  {journal} {Phys. Rev. D}\ }\textbf {\bibinfo {volume} {99}},\ \bibinfo
  {pages} {084030} (\bibinfo {year} {2019})},\ \Eprint
  {https://arxiv.org/abs/1811.07500} {arXiv:1811.07500 [gr-qc]} \BibitemShut
  {NoStop}%
\bibitem [{\citenamefont {Daka}\ \emph {et~al.}(2019)\citenamefont {Daka},
  \citenamefont {Phan},\ and\ \citenamefont {Kain}}]{Daka:2019iix}%
  \BibitemOpen
  \bibfield  {author} {\bibinfo {author} {\bibfnamefont {E.}~\bibnamefont
  {Daka}}, \bibinfo {author} {\bibfnamefont {N.~N.}\ \bibnamefont {Phan}},\
  and\ \bibinfo {author} {\bibfnamefont {B.}~\bibnamefont {Kain}},\ }\bibfield
  {title} {\bibinfo {title} {{Perturbing the ground state of Dirac stars}},\
  }\href {https://doi.org/10.1103/PhysRevD.100.084042} {\bibfield  {journal}
  {\bibinfo  {journal} {Phys. Rev. D}\ }\textbf {\bibinfo {volume} {100}},\
  \bibinfo {pages} {084042} (\bibinfo {year} {2019})},\ \Eprint
  {https://arxiv.org/abs/1910.09415} {arXiv:1910.09415 [gr-qc]} \BibitemShut
  {NoStop}%
\bibitem [{\citenamefont {Kain}(2023{\natexlab{c}})}]{Kain:2023jgu}%
  \BibitemOpen
  \bibfield  {author} {\bibinfo {author} {\bibfnamefont {B.}~\bibnamefont
  {Kain}},\ }\bibfield  {title} {\bibinfo {title} {{Einstein-Dirac system in
  semiclassical gravity}},\ }\href
  {https://doi.org/10.1103/PhysRevD.107.124001} {\bibfield  {journal} {\bibinfo
   {journal} {Phys. Rev. D}\ }\textbf {\bibinfo {volume} {107}},\ \bibinfo
  {pages} {124001} (\bibinfo {year} {2023}{\natexlab{c}})},\ \Eprint
  {https://arxiv.org/abs/2304.10627} {arXiv:2304.10627 [gr-qc]} \BibitemShut
  {NoStop}%
\bibitem [{\citenamefont {Calhoun}\ \emph {et~al.}(2022)\citenamefont
  {Calhoun}, \citenamefont {Fay},\ and\ \citenamefont
  {Kain}}]{Calhoun:2022xrw}%
  \BibitemOpen
  \bibfield  {author} {\bibinfo {author} {\bibfnamefont {K.}~\bibnamefont
  {Calhoun}}, \bibinfo {author} {\bibfnamefont {B.}~\bibnamefont {Fay}},\ and\
  \bibinfo {author} {\bibfnamefont {B.}~\bibnamefont {Kain}},\ }\bibfield
  {title} {\bibinfo {title} {{Matter traveling through a wormhole}},\ }\href
  {https://doi.org/10.1103/PhysRevD.106.104054} {\bibfield  {journal} {\bibinfo
   {journal} {Phys. Rev. D}\ }\textbf {\bibinfo {volume} {106}},\ \bibinfo
  {pages} {104054} (\bibinfo {year} {2022})},\ \Eprint
  {https://arxiv.org/abs/2210.04905} {arXiv:2210.04905 [gr-qc]} \BibitemShut
  {NoStop}%
\bibitem [{\citenamefont {Sakurai}(1967)}]{Sakurai}%
  \BibitemOpen
  \bibfield  {author} {\bibinfo {author} {\bibfnamefont {J.~J.}\ \bibnamefont
  {Sakurai}},\ }\href@noop {} {\emph {\bibinfo {title} {{Advanced Quantum
  Mechanics}}}}\ (\bibinfo  {publisher} {Addison-Wesley},\ \bibinfo {year}
  {1967})\BibitemShut {NoStop}%
\bibitem [{\citenamefont {Bransden}\ and\ \citenamefont
  {Joachain}(2000)}]{Bransden}%
  \BibitemOpen
  \bibfield  {author} {\bibinfo {author} {\bibfnamefont {B.~H.}\ \bibnamefont
  {Bransden}}\ and\ \bibinfo {author} {\bibfnamefont {C.~J.}\ \bibnamefont
  {Joachain}},\ }\href@noop {} {\emph {\bibinfo {title} {{Quantum
  Mechanics}}}},\ \bibinfo {edition} {2nd}\ ed.\ (\bibinfo  {publisher}
  {Prentice Hall},\ \bibinfo {year} {2000})\BibitemShut {NoStop}%
\bibitem [{\citenamefont {Birrell}\ and\ \citenamefont
  {Davies}(1984)}]{Birrell:1982ix}%
  \BibitemOpen
  \bibfield  {author} {\bibinfo {author} {\bibfnamefont {N.~D.}\ \bibnamefont
  {Birrell}}\ and\ \bibinfo {author} {\bibfnamefont {P.~C.~W.}\ \bibnamefont
  {Davies}},\ }\href@noop {} {\emph {\bibinfo {title} {{Quantum Fields in
  Curved Space}}}}\ (\bibinfo  {publisher} {Cambridge Univ. Press},\ \bibinfo
  {year} {1984})\BibitemShut {NoStop}%
\bibitem [{\citenamefont {Carroll}(2004)}]{Carroll:2004st}%
  \BibitemOpen
  \bibfield  {author} {\bibinfo {author} {\bibfnamefont {S.~M.}\ \bibnamefont
  {Carroll}},\ }\href@noop {} {\emph {\bibinfo {title} {{Spacetime and
  geometry: An introduction to general relativity}}}}\ (\bibinfo  {publisher}
  {Addison-Wesley},\ \bibinfo {year} {2004})\BibitemShut {NoStop}%
\bibitem [{\citenamefont {Wald}(1995)}]{Wald:1995yp}%
  \BibitemOpen
  \bibfield  {author} {\bibinfo {author} {\bibfnamefont {R.~M.}\ \bibnamefont
  {Wald}},\ }\href@noop {} {\emph {\bibinfo {title} {{Quantum Field Theory in
  Curved Space-Time and Black Hole Thermodynamics}}}},\ Chicago Lectures in
  Physics\ (\bibinfo  {publisher} {University of Chicago Press},\ \bibinfo
  {year} {1995})\BibitemShut {NoStop}%
\bibitem [{\citenamefont {Parker}\ and\ \citenamefont
  {Toms}(2009)}]{Parker:2009uva}%
  \BibitemOpen
  \bibfield  {author} {\bibinfo {author} {\bibfnamefont {L.~E.}\ \bibnamefont
  {Parker}}\ and\ \bibinfo {author} {\bibfnamefont {D.}~\bibnamefont {Toms}},\
  }\href {https://doi.org/10.1017/CBO9780511813924} {\emph {\bibinfo {title}
  {{Quantum Field Theory in Curved Spacetime}: {Quantized Field and
  Gravity}}}},\ Cambridge Monographs on Mathematical Physics\ (\bibinfo
  {publisher} {Cambridge University Press},\ \bibinfo {year}
  {2009})\BibitemShut {NoStop}%
\bibitem [{\citenamefont {Hu}\ and\ \citenamefont
  {Verdaguer}(2020)}]{Hu:2020luk}%
  \BibitemOpen
  \bibfield  {author} {\bibinfo {author} {\bibfnamefont {B.-L.~B.}\
  \bibnamefont {Hu}}\ and\ \bibinfo {author} {\bibfnamefont {E.}~\bibnamefont
  {Verdaguer}},\ }\href {https://doi.org/10.1017/9780511667497} {\emph
  {\bibinfo {title} {{Semiclassical and Stochastic Gravity}: {Quantum Field
  Effects on Curved Spacetime}}}},\ Cambridge Monographs on Mathematical
  Physics\ (\bibinfo  {publisher} {Cambridge University Press},\ \bibinfo
  {year} {2020})\BibitemShut {NoStop}%
\bibitem [{\citenamefont {Alcubierre}\ \emph {et~al.}(2023)\citenamefont
  {Alcubierre}, \citenamefont {Barranco}, \citenamefont {Bernal}, \citenamefont
  {Degollado}, \citenamefont {Diez-Tejedor}, \citenamefont {Megevand},
  \citenamefont {N\'u\~nez},\ and\ \citenamefont
  {Sarbach}}]{Alcubierre:2022rgp}%
  \BibitemOpen
  \bibfield  {author} {\bibinfo {author} {\bibfnamefont {M.}~\bibnamefont
  {Alcubierre}}, \bibinfo {author} {\bibfnamefont {J.}~\bibnamefont
  {Barranco}}, \bibinfo {author} {\bibfnamefont {A.}~\bibnamefont {Bernal}},
  \bibinfo {author} {\bibfnamefont {J.~C.}\ \bibnamefont {Degollado}}, \bibinfo
  {author} {\bibfnamefont {A.}~\bibnamefont {Diez-Tejedor}}, \bibinfo {author}
  {\bibfnamefont {M.}~\bibnamefont {Megevand}}, \bibinfo {author}
  {\bibfnamefont {D.}~\bibnamefont {N\'u\~nez}},\ and\ \bibinfo {author}
  {\bibfnamefont {O.}~\bibnamefont {Sarbach}},\ }\bibfield  {title} {\bibinfo
  {title} {{Boson stars and their relatives in semiclassical gravity}},\ }\href
  {https://doi.org/10.1103/PhysRevD.107.045017} {\bibfield  {journal} {\bibinfo
   {journal} {Phys. Rev. D}\ }\textbf {\bibinfo {volume} {107}},\ \bibinfo
  {pages} {045017} (\bibinfo {year} {2023})},\ \Eprint
  {https://arxiv.org/abs/2212.02530} {arXiv:2212.02530 [gr-qc]} \BibitemShut
  {NoStop}%
\bibitem [{\citenamefont {Finster}\ \emph
  {et~al.}(1999{\natexlab{c}})\citenamefont {Finster}, \citenamefont
  {Smoller},\ and\ \citenamefont {Yau}}]{Finster:1998ju}%
  \BibitemOpen
  \bibfield  {author} {\bibinfo {author} {\bibfnamefont {F.}~\bibnamefont
  {Finster}}, \bibinfo {author} {\bibfnamefont {J.}~\bibnamefont {Smoller}},\
  and\ \bibinfo {author} {\bibfnamefont {S.-T.}\ \bibnamefont {Yau}},\
  }\bibfield  {title} {\bibinfo {title} {{Nonexistence of black hole solutions
  for a spherically symmetric, static Einstein-Dirac-Maxwell system}},\ }\href
  {https://doi.org/10.1007/s002200050675} {\bibfield  {journal} {\bibinfo
  {journal} {Commun. Math. Phys.}\ }\textbf {\bibinfo {volume} {205}},\
  \bibinfo {pages} {249} (\bibinfo {year} {1999}{\natexlab{c}})},\ \Eprint
  {https://arxiv.org/abs/gr-qc/9810048} {arXiv:gr-qc/9810048} \BibitemShut
  {NoStop}%
\bibitem [{\citenamefont {Olabarrieta}\ \emph {et~al.}(2007)\citenamefont
  {Olabarrieta}, \citenamefont {Ventrella}, \citenamefont {Choptuik},\ and\
  \citenamefont {Unruh}}]{Olabarrieta:2007di}%
  \BibitemOpen
  \bibfield  {author} {\bibinfo {author} {\bibfnamefont {I.}~\bibnamefont
  {Olabarrieta}}, \bibinfo {author} {\bibfnamefont {J.~F.}\ \bibnamefont
  {Ventrella}}, \bibinfo {author} {\bibfnamefont {M.~W.}\ \bibnamefont
  {Choptuik}},\ and\ \bibinfo {author} {\bibfnamefont {W.~G.}\ \bibnamefont
  {Unruh}},\ }\bibfield  {title} {\bibinfo {title} {{Critical Behavior in the
  Gravitational Collapse of a Scalar Field with Angular Momentum in Spherical
  Symmetry}},\ }\href {https://doi.org/10.1103/PhysRevD.76.124014} {\bibfield
  {journal} {\bibinfo  {journal} {Phys. Rev. D}\ }\textbf {\bibinfo {volume}
  {76}},\ \bibinfo {pages} {124014} (\bibinfo {year} {2007})},\ \Eprint
  {https://arxiv.org/abs/0708.0513} {arXiv:0708.0513 [gr-qc]} \BibitemShut
  {NoStop}%
\bibitem [{\citenamefont {Morris}\ and\ \citenamefont
  {Thorne}(1988)}]{Morris:1988cz}%
  \BibitemOpen
  \bibfield  {author} {\bibinfo {author} {\bibfnamefont {M.~S.}\ \bibnamefont
  {Morris}}\ and\ \bibinfo {author} {\bibfnamefont {K.~S.}\ \bibnamefont
  {Thorne}},\ }\bibfield  {title} {\bibinfo {title} {{Wormholes in space-time
  and their use for interstellar travel: A tool for teaching general
  relativity}},\ }\href {https://doi.org/10.1119/1.15620} {\bibfield  {journal}
  {\bibinfo  {journal} {Am. J. Phys.}\ }\textbf {\bibinfo {volume} {56}},\
  \bibinfo {pages} {395} (\bibinfo {year} {1988})}\BibitemShut {NoStop}%
\end{thebibliography}

%

\end{document}